\documentclass{article}

 \usepackage[preprint]{neurips_2026}

\usepackage[utf8]{inputenc} %
\usepackage[T1]{fontenc}    %
\usepackage{hyperref}       %
\usepackage{url}            %
\usepackage{booktabs}       %
\usepackage{amsfonts}       %
\usepackage{nicefrac}       %
\usepackage{microtype}      %
\usepackage[table]{xcolor}         %
\usepackage{amsmath}
\usepackage{graphicx}
\usepackage{wrapfig}
\usepackage{placeins}
\usepackage{needspace}
\usepackage{enumitem}

\usepackage{amsthm}
\newtheorem{proposition}{Proposition}
\newtheorem{condition}{Condition}
\title{Attribution-Guided Shared-Private Decoupling for Noise-Reduced Audio-Visual Representation Learning}

\author{
Linge Wang$^{1,2}$ \quad
Yingying Chen$^{1,2,\dagger}$ \quad
Bingke Zhu$^{1,2}$ \quad
Lu Zhou$^{1,2}$\\
\textbf{Haixin Wang}$^{1,2}$ \quad
\textbf{Zhipiao Liu}$^{4}$ \quad
\textbf{Jinqiao Wang}$^{1,2,3,\dagger}$\\[3pt]
$^{1}$ Foundation Model Research Center, Institute of Automation,\\
Chinese Academy of Sciences, Beijing, China\\
$^{2}$ School of Artificial Intelligence, University of Chinese Academy of Sciences,\\
Beijing, China\\
$^{3}$ Objecteye Inc., Beijing, China
$^{4}$ Huawei Technologies Ltd., Beijing, China\\[2pt]
{\ttfamily\small wanglinge2024@ia.ac.cn}\\
{\ttfamily\small
\{yingying.chen,bingke.zhu,lu.zhou,jqwang\}@nlpr.ia.ac.cn}\\
{\small $^\dagger$Corresponding authors}
}

\begin{document}

\maketitle
\begin{abstract}

Audio-visual representation learning commonly aligns paired audio and visual
signals with sample-level contrastive objectives. However, real-world
audio-visual pairs usually contain heterogeneous patch tokens: only a subset carries
reliable cross-modal shared semantics, while many others mainly encode
modality-private residuals. Aggregating all tokens into a single global
representation can therefore introduce semantic noise into cross-modal
alignment. Motivated by the minimal-sufficiency principle, we propose an
attribution-guided shared-private decoupling framework to mitigate this issue.
Our key idea is to identify
shared-cover subviews that preserve dominant cross-modal information while
suppressing modality-private residuals, and use them to supervise decoupled
shared and private representations. To identify such subviews efficiently, we draw on transformer attribution method and introduce Layer-Truncated Attribution (LTA), which provides a lightweight estimate of token-level relevance to cross-modal shared information. We further adopt a teacher-student framework, where the teacher provides attribution-induced supervision and the student internalizes shared-private decoupling into its representations, without introducing extra inference cost. Experiments on retrieval, classification, and sound-prompted semantic
segmentation show consistent gains over strong baselines. Compared with baseline,
our method improves the average zero-shot R@1 by 4.0 across
AudioSet and VGGSound retrieval, and improves AS20K classification mAP by
3.1, demonstrating the effectiveness of fine grained attribution guidance for
reducing semantic noise. 
Code will be released at \url{https://github.com/wanglg20/LTA}.
\end{abstract}
\section{Introduction}

Multimodal representation learning aims to learn transferable representations by exploiting the complementary information across different sensory modalities. 
Early studies have shown that jointly modeling multiple modalities can improve feature learning, cross-modal retrieval, and semantic grounding~\cite{ngiam2011multimodal,srivastava2012multimodal,frome2013devise,kiros2014unifying,lu2019vilbert,radford2021learning}. 
Among various multimodal settings, audio-visual representation learning is particularly important because real world videos naturally provide massive paired audio and visual signals, where sounds are often correlated with visible objects, events, and temporal dynamics. 
This has motivated a series of self-supervised audio-visual learning methods that exploit audio-video correspondence from unlabeled videos~\cite{aytar2016soundnet, huang2023mavil,girdhar2023imagebind,georgescu2023audiovisual,guo2024crossmae}. 
More recently, representative frameworks ~\cite{gong2022cavmae,araujo2025cav,linge26} combine masked autoencoding~\cite{he2022masked} with contrastive audio-visual alignment based on InfoNCE~\cite{oord2018representation}, leading to strong transferable representations for downstream tasks such as cross-modal retrieval, event recognition, and sound source localization.

Despite these advances, real-world audio-visual pairs are rarely perfectly aligned. 
From the perspective of partial information decomposition, paired modalities generally contain both shared information that is redundant across modalities and unique information that is specific to each modality~\cite{williams2010nonnegative,bertschinger2014quantifying,Dufumier2025PID}. 
In audio-visual data, the shared component is often reflected in acoustically
grounded visual events, such as a speaking person, a musical instrument being
played, or an object action synchronized with sound. In contrast,
modality-private components may correspond to off-screen sounds, silent visual
regions, scene context, or appearance details that have no direct counterpart
in the other modality.
Most existing audio-visual contrastive frameworks, however, encode each modality into a single global token and align audio and visual tokens with a sample-level InfoNCE objective. 
This design has two inherent limitations. 
First, sample-level contrastive learning mainly supervises the cross-modal shared component, since the correspondence between paired audio and visual signals is induced by their shared semantics. As a result, modality-private cues, although potentially useful for downstream recognition, are not explicitly captured and may instead be absorbed into the same global representation.
Second, the global token is obtained by aggregating all patch tokens, although these tokens are semantically heterogeneous, only a subset carries reliable cross-modal correspondence, while many others are dominated by modality-private residuals. 
As a result, low-relevance patches can be inadvertently involved in cross-modal alignment, introducing semantic noise into the learned representation. 
Recent studies have explored decoupled or disentangled multimodal representations~\cite{Dufumier2025PID,gao2025disentangled,liuplug,xia2023achieving}, but they mainly focus on modality-level or representation-level decomposition and do not explicitly address how shared and private information should be separated at the fine-grained patch token level.

To address these issues, we take inspiration from the minimal-sufficiency principle of the information bottleneck~\cite{tishby2000ori_IB, DeepIB} and seek a representation that preserves the shared information while suppressing modality-private residuals. 
Our key idea is to construct a shared-cover subview, a subset of patch tokens that retains sufficient cross-modal shared information but filters out tokens dominated by modality-private information. 
Based on this subview, we decouple each modality representation into two learnable tokens, a shared token $z_s^m$ for cross-modal alignment and a private token $z_p^m$ for modality-private residual information. 
The main challenge is how to identify such shared-relevant tokens without manual annotations. 
To this end, we build on Transformer attribution methods\cite{achtibat24AttnLRP, Bousselham2024LeGrad, chefer2021transformer, Chefer2021GenericAE} and introduce LTA, which estimates token-level relevance from the gradient of the cross-modal matching score with respect to the last layer attention map. 
LTA provides an efficient and objective-conditioned ranking signal for selecting shared-cover subviews. 
Finally, we instantiate the framework with a teacher-student training scheme. The teacher produces attribution-induced high-relevance and low-relevance subview anchors, and the student learns to associate its shared and private tokens with the corresponding anchors. 
After training, the student internalizes the attribution-guided decoupling behavior and requires no extra computation during inference.

In summary, our main contributions are as follows:
\begin{itemize}[leftmargin=1.2em, labelsep=0.35em]
     \item We develop an attribution-guided teacher-student framework for shared-private decoupling, where two learnable tokens are trained to capture cross-modal shared information and modality-private residuals through high-relevance and low-relevance subview supervision.

    \item We introduce Layer-Truncated Attribution (LTA), an efficient token-level relevance estimator that uses objective conditioned gradient-attention signals from the last Transformer block to identify patch tokens carrying cross-modal shared information without full relevance propagation.

    \item Extensive experiments on large scale real world audio-visual data show state-of-the-art performance on retrieval, classification, and sound-prompted semantic segmentation, demonstrating the effectiveness of fine-grained attribution guidance for reducing semantic noise.
\end{itemize}
\section{Related Works}
\paragraph{Audio-Visual Representation Learning.}
Audio-visual representation learning has long explored the natural correspondence between sound and vision, from early statistical modeling of audio-visual association~\cite{statistical, Probabalistic} to deep self-supervised learning from unlabeled videos~\cite{aytar2016soundnet, owens2016ambientsoundprovidessupervision, NEURIPS2020_6f2268bd}. 
A major line of work learns cross-modal correspondence by discriminating synchronized or paired audio-visual samples~\cite{arandjelovic2017look, 10.5555/3327757.3327874, morgado2021crossmodal}, which also enables weakly supervised localization of sounding objects and events~\cite{arandjelovic2018objects, Lee2021CrossAttentionalAF}. 
More recently, masked modeling has become a dominant paradigm for audio-visual pretraining, reconstructing masked audio and visual tokens or exploiting cross-modal prediction~\cite{huang2023mavil, georgescu2023audiovisual, guo2024crossmae}. 
Representative methods further combine masked autoencoding with contrastive alignment, achieving strong transferable representations for retrieval, classification, and localization~\cite{gong2022cavmae, araujo2025cav, linge26}. 
Despite these advances, most existing methods rely on sample-level objectives and aggregate heterogeneous patch tokens into a single global representation, leaving fine-grained token-level information decoupling underexplored.
Our work takes a step toward this direction by studying how audio-visual representations can be separated at the token level to reduce the influence of semantically noisy regions.

\paragraph{Transformer Attribution.}
Model attribution aims to explain neural predictions by identifying input regions or features that contribute most to a target output~\cite{molnar2020interpretable, ribeiro2016should}. 
Early studies developed feature visualization and gradient- or relevance-based explanations for deep networks~\cite{erhan2009visualizing, zeiler2014visualizing, DTD, sundararajan2017axiomatic, selvaraju2017grad}. 
For Transformers, raw attention weights are often insufficient as faithful explanations~\cite{jain-wallace-2019-attention, serrano-smith-2019-attention}, motivating attention propagation methods such as attention rollout and attention flow~\cite{abnar-zuidema-2020-quantifying}. 
Subsequent works further combine attention with gradients or relevance propagation to obtain class- or objective-conditioned explanations for vision and multimodal Transformers~\cite{chefer2021transformer, Chefer2021GenericAE, Hao_Dong_Wei_Xu_2021}. 
Recent methods improve this line by enhancing computational efficiency or providing more principled relevance propagation for Transformer architectures~\cite{chen2023beyond, Bousselham2024LeGrad, achtibat24AttnLRP}.
\section{Theoretical Motivation}
\subsection{Problem Setup}
Considering a paired audio-visual sample $X=(A,V)$, existing audio-visual representation learning frameworks typically process the two modalities with modality-specific encoders and then employ a unified block to aggregate token-level representations before applying a cross-modal alignment objective~\cite{araujo2025cav, gong2022cavmae}. 
For each modality, the input can be viewed as a sequence of patch tokens. 
We write the audio and visual inputs as
\begin{equation}
    A=\{A_i\}_{i=1}^{N_A}, 
    \qquad 
    V=\{V_j\}_{j=1}^{N_V},
\end{equation}
where $A_i$ denotes the $i$-th audio patch token and $V_j$ denotes the $j$-th visual patch token.

Let $Y$ denote the downstream task variable. 
For an audio-visual pair $(A,V)$, the task relevant information may arise from both cross-modal shared semantics and modality-specific cues. 

Conceptually, the information contained in each modality can be attributed to two latent sources: shared information and modality-private residual information. 
We write this shared-private abstraction as
\begin{equation}
    A \leftarrow (S, P_A), 
    \qquad 
    V \leftarrow (S, P_V), 
    \qquad 
    Y \leftarrow (S, P_A, P_V),
    \label{eq:decomp}
\end{equation}
where $S$ denotes the information shared across modalities, and $P_A$ and $P_V$ denote audio-private and visual-private residual information, respectively. 

Representing each modality with a single global token tends to mix cross-modal shared information with modality-private residuals. 
Following the shared-private abstraction in Eq.~\eqref{eq:decomp}, we instead represent each modality with two learnable tokens: a shared token \(z_s^m\) for \(S\), and a private token \(z_p^m\) for \(P_m\). 
The key challenge is that shared and modality-private components are entangled at the patch-token level.
We formalize this challenge next from a minimal-sufficiency perspective.

\subsection{Minimal-Sufficient Subviews for Shared Information Representation}
We first characterize the desired shared representation from the information bottleneck\cite{DeepIB} perspective. 
For modality $m\in\{A,V\}$, the shared token $z_s^m$ should preserve information about the cross-modal shared information $S$ while suppressing modality-private residual information $P_m$. 
We therefore use the following IB style criterion to describe an ideal shared representation:
\begin{equation}
    \max_{z_s^m}\; I(z_s^m;S)-\beta I(z_s^m;P_m),
    \label{eq:ib_shared_rep}
\end{equation}
where $\beta>0$ balances shared sufficiency and private minimality.
To instantiate this criterion, we consider the audio modality as an example.
Let \(A=\{A_i\}_{i=1}^{N_A}\) denote the audio patch tokens.
Since shared and modality-private information is entangled at the token level, the full token sequence may contain redundant private residuals under the minimal-sufficiency criterion.
We therefore seek a shared-cover subview that preserves the dominant information about \(S\) while filtering out tokens with low shared relevance.
A natural construction is to rank tokens according to their contribution to the shared information \(S\).
We introduce an ideal shared-relevance score:
\begin{equation}
    r_i^S = \mathcal{R}(A_i; S),
\end{equation}
where $r_i^S$ measures the contribution of token $A_i$ to the shared information $S$. 

Sorting tokens by $r_i^S$ in descending order gives a permutation $\pi$ such that
\begin{equation}
    r_{\pi_1}^S \ge r_{\pi_2}^S \ge \cdots \ge r_{\pi_{N_A}}^S .
\end{equation}
Given a retention ratio $\rho\in(0,1]$, we keep the top $K=\lfloor \rho N_A \rfloor$ tokens as the selected subview:
\begin{equation}
    A_{\mathcal{T}}=\{A_{\pi_i}\}_{i=1}^{K},
    \qquad
    A_{\mathcal{R}}=A\setminus A_{\mathcal{T}},
\end{equation}
where $A_{\mathcal{R}}$ denotes the discarded complement.
For this score-induced subview to improve the minimal-sufficiency objective, we characterize its expected behavior through two sufficient conditions.

\begin{condition}[Shared Cover]
\label{cond:shared_cover}
Let $A_{\mathcal{T}}$ be the selected shared-cover subview and 
$A_{\mathcal{R}} = A \setminus A_{\mathcal{T}}$ be its discarded complement. 
There exists $\epsilon_S \ge 0$ such that
\begin{equation}
    I(A_{\mathcal{R}}; S \mid A_{\mathcal{T}}) \le \epsilon_S .
    \label{eq:shared_cover_condition}
\end{equation}
\end{condition}
This condition captures the desired effect of shared-relevance selection: the retained subview should preserve most information about the shared information $S$, leaving only limited additional shared information in the discarded complement.

\begin{condition}[Private Tail]
\label{cond:private_tail}
For the decomposition $A = A_{\mathcal{T}} \cup A_{\mathcal{R}}$, 
there exists $\epsilon_P > 0$ such that
\begin{equation}
    I(A_{\mathcal{R}}; P_A \mid A_{\mathcal{T}}) \ge \epsilon_P .
    \label{eq:private_tail_condition}
\end{equation}
\end{condition}
where $\epsilon_P>0$.
This condition reflects that low-shared-relevance tokens are not necessarily information-free.
Although they contribute little additional information about the shared information $S$, they may still encode audio-private residual information.

For any token subset $B\subseteq A$, we define its input-level IB criterion as
\begin{equation}
    F(B)=I(B;S)-\beta I(B;P_A),
    \label{eq:subview_ib_objective}
\end{equation}

\begin{proposition}[Shared-Cover Subview Improves the IB Objective]
\label{prop:shared_cover_ib}
Under the Shared Cover and Private Tail conditions, the selected subview $A_{\mathcal{T}}$ satisfies
\begin{equation}
    F(A_{\mathcal{T}}) - F(A) \geq \beta\epsilon_P - \epsilon_S .
\end{equation}
In particular, if $\beta\epsilon_P > \epsilon_S$, then $F(A_{\mathcal{T}}) > F(A)$.
\end{proposition}
This proposition shows that selecting a shared-cover subview can be more suitable than using the full input under the minimal-sufficiency objective. The proof is provided in Appendix~\ref{app:proof_subview_ib}.
\section{Proposed Method}

\subsection{ Efficient Token-Relevance Estimation via Layer-Truncated Attribution}
The theoretical motivation above suggests that constructing a shared-cover subview relies on a reliable token-level scoring function: tokens with high scores should preserve the dominant shared factor, while tokens with low scores are more likely to contain modality-private residual information. 
Therefore, the key question is how to estimate the shared relevance of each token without introducing additional supervision or task-specific annotations.

A straightforward choice is to use raw attention weights as token importance scores. 
However, attention weights only describe the forward aggregation pattern and are not necessarily faithful to the cross-modal matching objective~\cite{serrano-smith-2019-attention, jain-wallace-2019-attention}. 
Attribution methods provide a more suitable alternative by measuring token contributions with respect to a specified output or objective, which matches our need to identify tokens that drive cross-modal matching. 
Inspired by CheferCAM~\cite{chefer2021transformer}, which combines attention maps with their objective gradients, we use gradient-attention relevance as an objective-conditioned token scoring signal:
\begin{equation}
    \mathrm{CAM}^{(l)}
    =
    I
    +
    \mathbb{E}_{h}
    \left[
    \left(
    \nabla A^{(l)}_{h} \odot A^{(l)}_{h}
    \right)^{+}
    \right],
    \label{eq:cam}
\end{equation}
where $A^{(l)}_{h}$ denotes the attention map of head $h$ at layer $l$, 
$\nabla A^{(l)}_{h}$ is the gradient of the cross-modal matching score with respect to $A^{(l)}_{h}$, 
$\mathbb{E}_{h}[\cdot]$ denotes averaging over attention heads, 
$\odot$ denotes element-wise multiplication, and $(\cdot)^{+}$ keeps positive relevance contributions.

Standard Transformer attribution methods usually propagate relevance through the full network to obtain a complete input-level explanation. 
In contrast, our goal is to construct shared-cover subviews, which only requires a relative ranking of token contributions to the cross-modal matching objective. 
We therefore propose Layer-Truncated Attribution (LTA), which computes Eq.~\eqref{eq:cam} only at the last Transformer block, where the CLS token directly aggregates patch tokens for cross-modal matching. 
This preserves the objective conditioned nature of attribution while avoiding full network relevance propagation. 
Empirically, LTA scores provide a strong ranking signal for subview construction: 
the selected subviews substantially improve cross-modal retrieval over full-view inputs 
and other token scoring baselines in Sec.~\ref{support}, while LTA also produces 
CheferCAM-like localization patterns with only 6.54\% extra computation over a 
forward pass, as analyzed in Appendix~\ref{app:lta_analysis}.

Concretely, we instantiate LTA by measuring how each patch token contributes to the cross-modal matching score through the last layer CLS attention. Fig.~\ref{fig:pipeline}(b) provides an overview of the LTA computation.
For an audio-visual pair, let
$ s_{av} = \langle z_{\mathrm{cls}}^a, z_{\mathrm{cls}}^v \rangle$
be the cross-modal matching score. 
For modality $m\in\{a,v\}$, we define the relevance score of token $i$ as
\begin{equation}
    r_i^m
    =
    \mathbb{E}_{h}
    \left[
    \left(
    \frac{\partial s_{av}}
    {\partial A^{m,L}_{h,\mathrm{cls},i}}
    A^{m,L}_{h,\mathrm{cls},i}
    \right)^{+}
    \right],
    \label{eq:lta}
\end{equation}
where $A^{m,L}_{h,\mathrm{cls},i}$ denotes the attention weight from the CLS token to token $i$ at head $h$ of the last Transformer block, and $\mathbb{E}_{h}[\cdot]$ denotes averaging over attention heads.

\subsection{Training Framework}
\begin{figure*}[t]
    \centering
    \includegraphics[width=\textwidth]{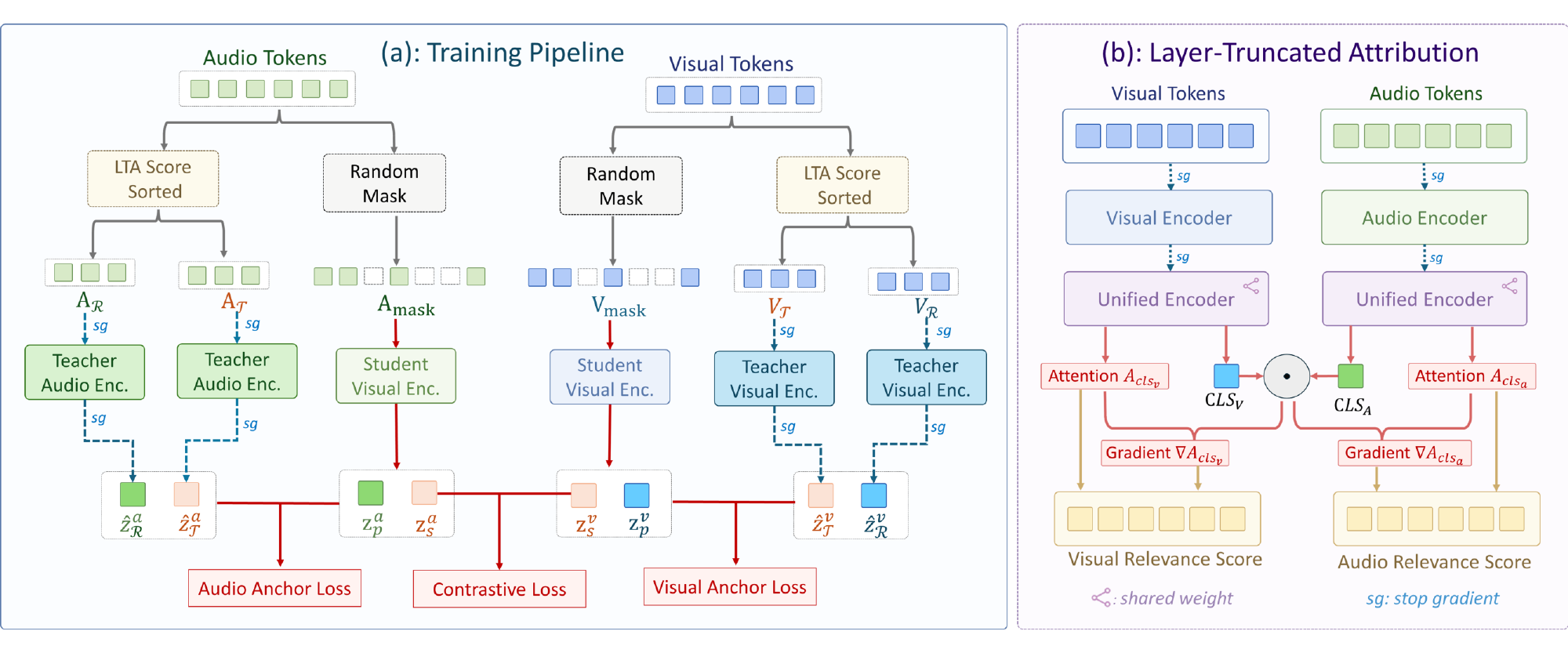}
    \caption{
Overall pipeline of our proposed framework.
(a) Attribution-guided teacher-student training. 
A pretrained teacher generates high-relevance shared-cover subviews and low-relevance residual complements, which serve as stop-gradient anchors for training the student's shared and private tokens. 
For clarity, we omit the MAE branch inherited from the baseline~\cite{linge26}. 
(b) Layer-Truncated Attribution (LTA). 
Given the cross-modal matching score, LTA computes CLS-to-token gradient-attention relevance from the last Transformer block. 
}
    \label{fig:pipeline}
\end{figure*}

The preceding sections establish two key ingredients: a shared-cover subview can reduce modality-private residuals at the input level, and LTA provides an efficient way to estimate such subviews. 
However, applying LTA at inference would require paired audio-visual inputs, gradient-based scoring, and explicit subview construction. 
We therefore adopt a teacher-student training framework, as shown in Fig.~\ref{fig:pipeline}(a), to distill the attribution-induced decoupling behavior into the student. 
The teacher generates high-relevance shared-cover targets and low-relevance residual complement targets, while the student learns to align its shared and private tokens with the corresponding responses. 
Once trained, the student internalizes this behavior and introduces no additional parameters or computational overhead during inference.

\subsubsection{Pretraining an Early Converged Relevance Teacher}

The teacher is pretrained from scratch following the baseline framework of~\cite{linge26}, which separates masked autoencoding (MAE) and contrastive learning (CL) into two forward passes and uses a lower masking ratio for the CL branch. 
At this stage, the teacher is trained with the original random masking strategy only, without attribution-guided subview selection. 
Rather than serving as a fully optimized representation model, its role is to provide token-relevance guidance for student training. 
As analyzed in Appendix~\ref{app:teacher_dynamic}, LTA-induced subviews become stable before the full training schedule is completed. 
Therefore, we stop teacher pretraining at epoch 18 instead of following the original 35-epoch TG-DP schedule.
\subsubsection{Student Training}

After teacher pretraining, we freeze the teacher and use it to generate
attribution-induced subview targets for each audio-visual pair. As shown in
Fig.~\ref{fig:pipeline}(a), for modality $m\in\{a,v\}$, the teacher computes
token relevance scores following Eq.~\eqref{eq:lta}, and separates
the input tokens into a high-relevance subview $X_{\mathcal{T}}^m$ and a
low-relevance complement $X_{\mathcal{R}}^m$. The frozen teacher then processes
these two subviews and produces stop-gradient anchor targets:
\begin{equation}
    \hat z_{\mathcal{T}}^m
    =
    \mathrm{sg}\left(f_{\theta_T}(X_{\mathcal{T}}^m)\right),
    \qquad
    \hat z_{\mathcal{R}}^m
    =
    \mathrm{sg}\left(f_{\theta_T}(X_{\mathcal{R}}^m)\right),
\end{equation}
where $\mathrm{sg}(\cdot)$ denotes stop-gradient.

The student follows the same backbone architecture, but replaces the original
single learnable global token with two learnable tokens for each modality: a
shared token and a private token. Their output representations are denoted as
$z_s^m$ and $z_p^m$, respectively. The shared token is encouraged to match the
teacher response on the shared-cover subview, while the private token is
encouraged to absorb the residual information from the low-relevance
complement:
\begin{equation}
    \mathcal{L}_{s\text{-anc}}
    =
    \sum_{m\in\{a,v\}}
    \ell_{\cos}\left(z_s^m, \hat z_{\mathcal{T}}^m\right),
    \qquad
    \mathcal{L}_{p\text{-anc}}
    =
    \sum_{m\in\{a,v\}}
    \ell_{\cos}\left(z_p^m, \hat z_{\mathcal{R}}^m\right),
\end{equation}
where $\ell_{\cos}(x,y)=1-\cos(x,y)$ denotes the cosine similarity loss.

To further stabilize representation learning under different masking ratios,
we introduce a consistency regularizer on the contrastive token. Specifically,
the CL branch with mask ratio $0.5$ produces the standard contrastive
representation $z_{\mathrm{cl},0.5}^m$, which is detached and used as the
target. We then reuse the intermediate features from the MAE branch and obtain
the corresponding CL representation under the higher mask ratio $0.75$, denoted
as $z_{\mathrm{cl},0.75}^m$. The consistency loss is defined as
\begin{equation}
    \mathcal{L}_{\mathrm{cons}}
    =
    \sum_{m\in\{a,v\}}
    \left\|
    z_{\mathrm{cl},0.75}^m
    -
    \mathrm{sg}\left(z_{\mathrm{cl},0.5}^m\right)
    \right\|_2^2 .
\end{equation}
This regularizer encourages the contrastive representation to remain stable
across different visible token ratios, while avoiding an additional full
forward pass by reusing the MAE branch computation.

The final student objective is
\begin{equation}
    \mathcal{L}_{\mathrm{student}}
    =
    \mathcal{L}_{\mathrm{base}}
    +
    \lambda_a (\mathcal{L}_{s\text{-anc}}
    + \mathcal{L}_{p\text{-anc}})
    +
    \lambda_c \mathcal{L}_{\mathrm{cons}},
\end{equation}
where $\mathcal{L}_{\mathrm{base}}$ denotes the original decoupled MAE and
contrastive training objective of~\cite{linge26}. In our framework, the
contrastive objective is applied only to the shared tokens. 
Through the attribution-induced anchor losses and the consistency
regularizer, the student learns to associate the shared token with shared-cover
subviews and the private token with residual complements, while maintaining
robust contrastive representations under different masking ratios. The resulting
model requires no additional computation during inference.
\section{Experiments}
\subsection{Experimental Setup}

Following standard practice in audio-visual representation
learning~\cite{gong2022cavmae, araujo2025cav, linge26}, we pretrain
our model on AudioSet-2M~\cite{gemmeke2017audio}. We evaluate the
pretrained representations on three downstream tasks: cross-modal
retrieval, audio-visual classification, and sound-prompted semantic
segmentation. Detailed dataset descriptions and training recipes
are provided in Appendix~\ref{app:implementation}. For VGGSound
retrieval, AVSBench localization, and AS20K classification, we report
means over repeated runs, with the repetition protocols and standard
deviations detailed in Appendix~\ref{app:repeated_runs}.
\paragraph{Computational cost.}
Under the same hardware setting of $8$ NVIDIA A100 GPUs, the overall
pretraining time increases from $19.0$ hours for the adopted backbone
to $27.1$ hours for our method. A stage-wise breakdown of the training
cost is provided in Appendix~\ref{app:lta_cost}.  Since the teacher and attribution computation are used only during training, our method introduces no extra computational overhead during inference.

\subsection{Main Results}

\paragraph{Cross-modal retrieval.}
We evaluate zero-shot audio-visual retrieval on the VGGSound and AudioSet-20K
evaluation sets provided by~\cite{araujo2025cav}. Following prior work, we
compute cosine similarity between frozen audio and visual representations and
report Recall@1, Recall@5, and Recall@10 in both audio-to-visual and
visual-to-audio directions. For our model, we use the shared token as the
global representation for retrieval, since this task primarily evaluates
cross-modal shared correspondence.

As shown in Table~\ref{tab:retrieval}, our method achieves state-of-the-art zero-shot retrieval performance on both AudioSet-20K and VGGSound\cite{chen2020vggsound}. Compared with the adopted backbone model TG-DP~\cite{linge26}, our method improves R@1 by 3.8 and 4.0 on AudioSet-20K for V$\rightarrow$A and A$\rightarrow$V retrieval, respectively. On VGGSound, it further improves R@1 by 4.0 and 3.7 in the two directions. These consistent gains demonstrate the effectiveness of our attribution guided shared-private decoupling framework for cross-modal alignment.
\begin{table*}[!ht]
\centering
\setlength{\tabcolsep}{5.2pt}
\resizebox{\linewidth}{!}{
\begin{tabular}{@{}l ccc ccc c ccc ccc@{}}
\toprule
& \multicolumn{6}{c}{\textbf{Vision $\rightarrow$ Audio}} & & \multicolumn{6}{c}{\textbf{Audio $\rightarrow$ Vision}}\\
\cmidrule(lr){2-7}\cmidrule(lr){9-14}
& \multicolumn{3}{c}{AudioSet Eval Subset} & \multicolumn{3}{c}{VGGSound Eval Subset} & & \multicolumn{3}{c}{AudioSet Eval Subset} & \multicolumn{3}{c}{VGGSound Eval Subset}\\
\cmidrule(lr){2-4}\cmidrule(lr){5-7}\cmidrule(lr){9-11}\cmidrule(lr){12-14}
\textbf{Method} & R@1 & R@5 & R@10 & R@1 & R@5 & R@10 & & R@1 & R@5 & R@10 & R@1 & R@5 & R@10\\
\midrule
CAV-MAE \cite{gong2022cavmae} & 16.1 & 38.6 & 49.3 & 14.7 & 35.3 & 45.9 & & 9.5 & 22.6 & 32.4 & 8.3 & 23.8 & 32.4\\
CAV-MAE\textsuperscript{Scale+++} \cite{araujo2025cav} & 18.8 & 39.5 & 50.1 & 14.8 & 34.2 & 44.0 & & 15.1 & 34.0 & 43.0 & 12.8 & 30.4 & 40.3\\
LanguageBind \cite{zhu2024languagebind} & 6.4 & 20.2 & 28.3 & 10.3 & 30.1 & 39.7 & & 4.4 & 15.0 & 22.5 & 6.5 & 22.7 & 33.5\\
AVSiam \cite{lin2024avsiam} & 19.7 & -- & -- & 19.0 & -- & -- & & 17.6 & -- & -- & 20.4 & -- & --\\
ImageBind \cite{girdhar2023imagebind} & 22.1 & 43.2 & 52.6 & 21.6 & 43.4 & 52.9 & & 20.8 & 42.6 & 51.6 & 20.7 & 43.2 & 53.4\\
DenseAV \cite{hamilton2024denseav}  & 26.3 & 49.9 & 58.5 & 24.2 & 48.2 & 58.3 & & 20.8 & 42.6 & 51.6 &  25.1 & 50.1 & 59.1\\

CAV\mbox{-}MAE Sync \cite{araujo2025cav} & 35.2 & 58.3 & 67.6 & 27.9 & 51.7 & 61.8 & & 27.9 & 52.4 & 62.2 & 23.2 & 46.2 & 58.1\\
TG-DP\cite{linge26} & 37.4 & 59.2 & 68.1 & 31.3 & 54.4 & 63.9 & & 37.1 & 59.7 & 68.2 & 30.3 & 53.1 & 64.2\\
\midrule
\rowcolor{gray!15} \textbf{Ours} & \textbf{41.2} & \textbf{62.8} & \textbf{70.8} & \textbf{35.3} & \textbf{59.8} & \textbf{69.5} & & \textbf{41.1} & \textbf{63.3} & \textbf{71.3} & \textbf{34.0} & \textbf{58.9} & \textbf{69.5}\\
\bottomrule
\end{tabular}
}
\caption{
Zero-shot cross-modal retrieval on the AudioSet-20K and VGGSound evaluation subsets. 
We report Recall@1, Recall@5, and Recall@10 for both Vision$\rightarrow$Audio (V$\rightarrow$A) and Audio$\rightarrow$Vision (A$\rightarrow$V) retrieval using frozen representations. 
}
\label{tab:retrieval}
\end{table*}
\begin{table}[t]
\centering
\begin{minipage}[t]{0.43\linewidth}
\centering
\setlength{\tabcolsep}{4pt}
\renewcommand{\arraystretch}{1.05}
\scriptsize
\resizebox{\linewidth}{!}{
\begin{tabular}{@{}lcc@{}}
\toprule
\textbf{Method} & \textbf{AS20K} & \textbf{VGGSound} \\
& \textbf{mAP} & \textbf{Acc.} \\
\midrule
CAV-MAE\textsuperscript{Scale+++}~\cite{araujo2025cav}  & 25.3 & 51.6 \\
MaViL~\cite{huang2023mavil} & 30.0 & -- \\
CAV-MAE Sync~\cite{araujo2025cav} & 30.5 & 52.7 \\
TG-DP~\cite{linge26} & 32.0 & 52.7 \\
\midrule
\rowcolor{gray!12}
\textbf{Ours} & \textbf{35.2} & \textbf{54.4} \\
\bottomrule
\end{tabular}
}
\vspace{0.35em}
\caption{
Audio-visual classification results under attention probing~\cite{araujo2025cav}.
}
\label{tab:classification}
\end{minipage}
\hfill
\begin{minipage}[t]{0.5\linewidth}
\centering
\setlength{\tabcolsep}{4pt}
\renewcommand{\arraystretch}{1.05}
\scriptsize
\resizebox{\linewidth}{!}{
\begin{tabular}{@{}lcccc@{}}
\toprule
& \multicolumn{2}{c}{\textbf{AVSBench}} 
& \multicolumn{2}{c}{\textbf{ADE20K-Sound}} \\
\cmidrule(lr){2-3}\cmidrule(lr){4-5}
\textbf{Method} & \textbf{mIoU} & \textbf{mAP} & \textbf{mIoU} & \textbf{mAP} \\
\midrule
DAVENet~\cite{DAVENET}            & --   & --   & 16.8 & 17.0 \\
CAV-MAE~\cite{gong2022cavmae}     & 28.0 & 24.5 & 21.2 & 20.9 \\
CAV-MAE Sync~\cite{araujo2025cav} & 22.3 & 25.2 & 22.6 & 22.7 \\
TG-DP~\cite{linge26}              & 21.8 & 24.3 & 21.5 & 22.3 \\
\midrule
\rowcolor{gray!12}
\textbf{Ours} & \textbf{29.9} & \textbf{37.3} & \textbf{26.1} & \textbf{24.0} \\
\bottomrule
\end{tabular}
}
\vspace{0.35em}
\caption{
Sound-prompted semantic segmentation results on AVSBench and ADE20K-Sound.
}
\label{tab:segmentation}
\end{minipage}
\vspace{-0.5em}
\end{table}

\paragraph{Audio-visual classification.}
We evaluate the representation quality of the pretrained models through
\emph{attention probing}, following the evaluation protocol
of~\cite{araujo2025cav}. Specifically, we freeze the pretrained audio and
visual encoders and train a lightweight attention-based classification head on
AudioSet-20K and VGGSound. For our model, we average the shared and private
tokens as the final global token for classification, ensuring that all compared
downstream models have the same
computational cost. We report mAP on AudioSet-20K and
top-1 accuracy on VGGSound.

As shown in Table~\ref{tab:classification}, our method achieves state-of-the-art audio-visual classification performance on both AudioSet-20K and VGGSound. Compared with TG-DP~\cite{linge26}, the adopted backbone model of our framework, our method improves AS20K mAP from $32.0$ to $35.2$. On VGGSound, our method improves top-1 accuracy from $52.7$ to $54.4$. These results demonstrate that attribution guided shared-private decoupling improves not only cross-modal retrieval but also the transferability of the learned representations to downstream recognition tasks.

\paragraph{Sound-prompted semantic segmentation.}

We further evaluate whether the learned representations capture fine-grained audio-visual correspondence through sound-prompted semantic segmentation. Following the protocol of DenseAV~\cite{hamilton2024denseav}, we use the similarity heatmap between visual patch tokens and the  audio shared CLS token as the segmentation prediction. We evaluate on AVSBench~\cite{AvsBench} and ADE20K-Sound~\cite{ADE20k, hamilton2024denseav}, and report mean intersection over union (mIoU) and mean average precision (mAP). 
As shown in Table~\ref{tab:segmentation}, our method achieves
the best sound-prompted semantic segmentation performance on
both benchmarks. On AVSBench, it improves the adopted backbone
TG-DP from $21.8$ to $29.9$ in mIoU and from $24.3$ to $37.3$
in mAP, showing a substantial gain in fine-grained audio-visual
localization. On ADE20K-Sound, our method improves mIoU from
$21.5$ to $26.1$ and mAP from $22.3$ to $24.0$. These results
indicate that attribution-guided shared-private decoupling leads
to more discriminative patch-level representations.

\subsection{Empirical Validation of LTA-Induced Subviews}
\label{support}
\begin{wraptable}{r}{0.5\linewidth}
\vspace{-1.0em}
\centering
\setlength{\tabcolsep}{3.0pt}
\renewcommand{\arraystretch}{1.05}
\scriptsize
\resizebox{0.98\linewidth}{!}{
\begin{tabular}{@{}lccc ccc@{}}
\toprule
& \multicolumn{3}{c}{\textbf{V$\rightarrow$A}} 
& \multicolumn{3}{c}{\textbf{A$\rightarrow$V}} \\
\cmidrule(lr){2-4}\cmidrule(lr){5-7}
\textbf{Subview Score} & \textbf{R@1} & \textbf{R@5} & \textbf{R@10}
& \textbf{R@1} & \textbf{R@5} & \textbf{R@10} \\
\midrule
Full view    & 31.7 & 56.0 & 65.9 & 30.5 & 54.0 & 65.3 \\
Similarity   & 21.6 & 46.2 & 56.4 & 23.2 & 44.8 & 54.0 \\
Attention    & 28.5 & 54.9 & 64.2 & 28.7 & 52.7 & 63.7 \\
CheferCAM~\cite{chefer2021transformer}   & 37.2  &  61.6 & 70.6  & 37.5  & 60.9 & 68.3  \\
\midrule
\rowcolor{gray!12}
\textbf{LTA} & \textbf{44.1} & \textbf{68.5} & \textbf{77.3}
             & \textbf{41.2} & \textbf{67.2} & \textbf{75.8} \\
\bottomrule
\end{tabular}
}
\caption{
Empirical validation of LTA-induced subviews on VGGSound. 
We compare teacher retrieval performance using subviews constructed by different token scoring functions.
}
\label{tab:lta_subview_retrieval}
\vspace{-1.0em}
\end{wraptable}

To verify whether LTA can construct more reliable shared-cover
subviews, we evaluate the induced subviews directly with the
pretrained teacher model. We freeze the teacher, select visible
tokens according to different scoring functions under the same
retention ratio of $0.50$, and feed the resulting subviews back
into the teacher encoder. The output representations are then
evaluated by zero-shot cross-modal retrieval.

As shown in Table~\ref{tab:lta_subview_retrieval}, LTA-induced
subviews substantially improve teacher retrieval performance
over the original full view. On VGGSound, LTA improves R@1 from
$31.7$ to $44.1$ for V$\rightarrow$A retrieval and from $30.5$
to $41.2$ for A$\rightarrow$V retrieval. In contrast, subviews
constructed by token-level similarity or raw attention perform
worse than the full view, indicating that these scores do not
reliably capture cross-modal shared relevance. We also compare
with the original CheferCAM, which performs full relevance
propagation through the Transformer. Although CheferCAM improves
over the full view, it remains clearly behind LTA. This result
suggests that full relevance propagation is not necessarily
optimal for subview selection.

To further examine the separation induced by LTA, we perform
the same retrieval diagnostic using only the low-relevance
residual tokens. Their R@1 drops to $8.9$ for V$\rightarrow$A
and $8.4$ for A$\rightarrow$V, compared with $44.1$ and $41.2$
for the high-relevance subviews. This contrast shows that LTA
concentrates cross-modal retrieval cues in the selected subviews,
while leaving substantially weaker shared cues in the residual
complements.

\begin{table}[!htbp]
\centering
\small
\renewcommand{\arraystretch}{1.08}
\begin{tabular*}{\linewidth}{@{\extracolsep{\fill}}lcccccc@{}}
\toprule
& \multicolumn{2}{c}{\textbf{Similarity}}
& \multicolumn{2}{c}{\textbf{Attention}}
& \multicolumn{2}{c}{\textbf{LTA}} \\
\cmidrule(lr){2-3}\cmidrule(lr){4-5}\cmidrule(lr){6-7}
\textbf{Pair condition} & mIoU & mAP & mIoU & mAP & mIoU & mAP \\
\midrule
Matched    & 22.3 & 25.2 & 34.2 & 42.8 & 44.2 & 60.1 \\
Mismatched & 22.2 & 25.4 & 34.4 & 43.2 & 34.7 & 45.7 \\
\bottomrule
\end{tabular*}
\caption{Sound-source localization under matched and mismatched audio-video pairing on AVSBench.
For mismatched pairs, the audio is replaced while the visual input and its original sound-source annotation are retained.
Both mIoU and mAP are reported in percentage.}
\label{tab:lta_mismatched}
\end{table}

We further examine whether LTA responds to audio-video
correspondence through a mismatched-pair diagnostic on AVSBench.
For each visual sample, we replace its matched audio with a
mismatched audio clip and evaluate the resulting localization
map against the original visual sound-source annotation.
As shown in Table~\ref{tab:lta_mismatched}, mismatched pairing
reduces LTA localization performance by $9.5$ mIoU and $14.4$
mAP, whereas similarity and raw attention change by at most
$0.4$ points. These results indicate that LTA is sensitive to
the correspondence between the paired modalities, supporting
its use as a cross-modal relevance signal for subview selection.

\subsection{Verification of Shared-Private Token Decoupling}
\begin{wraptable}{r}{0.5\textwidth}
\vspace{-1.0em}
\centering
\small
\vspace{-0.5em}
\resizebox{\linewidth}{!}{
\begin{tabular}{llccc}
\toprule
Token & Direction & R@1 & R@5 & R@10 \\
\midrule
Private & V$\rightarrow$A & 29.5 & 53.9 & 63.9 \\
Private & A$\rightarrow$V & 28.2 & 52.9 & 63.8 \\
\midrule
Shared & V$\rightarrow$A & 35.4 &59.6& 70.5 \\
Shared & A$\rightarrow$V & 34.3 & 57.9 & 69.8 \\
\bottomrule
\end{tabular}
}
\caption{
VGGSound zero-shot retrieval using different tokens as the global representation.
}
\label{tab:decouple}
\vspace{-1.0em}
\end{wraptable}

We further examine whether the learned shared and private tokens are functionally decoupled in the final representation.
Since cross-modal retrieval directly depends on shared audio-visual information, we use retrieval performance to measure how much shared information each token captures.
As shown in Table~\ref{tab:decouple}, the shared token consistently outperforms the private token for zero-shot retrieval on VGGSound.
Specifically, the shared token achieves 35.4 R@1 for V$\rightarrow$A and 34.3 R@1 for A$\rightarrow$V retrieval, while the private token obtains 29.5 and 28.2 R@1, respectively.
This gap shows that the shared token carries stronger cross-modal correspondence than the private token, which is consistent with our decoupled design.
At the same time, the private token still achieves non-trivial retrieval performance, indicating that the two tokens are not completely separated in terms of shared information.
This is also consistent with our formulation, since the residual complement is selected by relatively lower shared relevance rather than by the absence of shared semantics.

\subsection{Ablation Studies}
\begin{table}[!ht]
\centering
\setlength{\tabcolsep}{5pt}
\renewcommand{\arraystretch}{1.1}
\small
\begin{tabular}{@{}lccc ccc@{}}
\toprule
\textbf{Setting} 
& \textbf{Anchor} 
& \textbf{Cons.} 
& \textbf{Dec.} 
& \textbf{Avg. R@1} 
& \textbf{AS20K mAP}
& \textbf{Seg. mAP} \\
\midrule
Backbone 
&  &  &  
& 30.8 & 32.0 & 24.3 \\
+ Anchor 
& \checkmark &  &  
& 33.0 & 32.2 & 28.3 \\
+ Consistency 
& \checkmark & \checkmark &  
& 34.3 & 32.0 & 26.3 \\
+ Decoupled tokens 
& \checkmark & \checkmark & \checkmark 
& \textbf{34.9} & \textbf{35.1} & \textbf{38.2} \\
\bottomrule
\end{tabular}
\vspace{0.3em}
\caption{
Incremental ablation of the main training components. 
Anchor, Cons., and Dec. denote attribution-induced anchor supervision, consistency regularization, and shared-private token decoupling, respectively. 
Avg. R@1 is averaged over the two retrieval directions on VGGSound. Seg. mAP denotes sound-prompted semantic segmentation mAP on AVSBench.
}
\label{tab:component_ablation}
\end{table}
\paragraph{Effect of training components.}
We conduct an incremental ablation to examine the contribution of the main components in our training framework. Starting from the adopted backbone, we progressively add attribution-induced anchor supervision, consistency regularization, and shared-private token decoupling. As shown in Table~\ref{tab:component_ablation}, anchor supervision  improves the average R@1 from 30.8 to 33.0, showing that attribution-induced subview anchors provide useful guidance for cross-modal alignment. Adding consistency regularization further improves retrieval performance, suggesting that the regularized objective stabilizes global representation learning. Finally, introducing decoupled shared and private tokens leads to the best overall performance across all metrics. In particular, it improves AS20K classification mAP from 32.0 to 35.1 and localization mAP from 26.3 to 38.2. These gains indicate that explicit shared-private decoupling not only transfers subview supervision to fine-grained patch-level correspondence, but also preserves modality-private residual information that benefits recognition tasks requiring complementary audio-visual cues.

\paragraph{Effect of retention ratio.}
\begin{wraptable}{r}{0.5\linewidth}
\vspace{-1.0em}
\centering
\setlength{\tabcolsep}{4pt}
\renewcommand{\arraystretch}{1.05}
\scriptsize
\resizebox{0.98\linewidth}{!}{
\begin{tabular}{@{}lccc@{}}
\toprule
\textbf{Retention} 
& \textbf{Avg. R@1} 
& \textbf{AS20K mAP} 
& \textbf{Seg. mAP} \\
\midrule
0.25 & 32.6 & 35.3 & \textbf{38.3} \\
0.35 & 33.9 & 35.2 & 37.3 \\
\rowcolor{gray!12}
0.50 & \textbf{34.9} & 35.1 & 38.2 \\
0.75 & 34.8 & \textbf{35.5} & 31.5 \\
\bottomrule
\end{tabular}
}
\caption{
Effect of the retention ratio. 
}
\label{tab:retention_ratio}
\vspace{-1.0em}
\end{wraptable}
We further study the retention ratio used for constructing attribution-induced
subviews. As shown in Table~\ref{tab:retention_ratio}, increasing the ratio
from $0.25$ to $0.50$ consistently improves retrieval, suggesting that overly
compact subviews may discard useful shared information. Further increasing the
ratio to $0.75$ brings little retrieval gain and noticeably degrades
localization, indicating that excessive token retention can reintroduce
modality-private or weakly related regions. Classification remains relatively
stable across ratios, showing that global recognition is less sensitive to this
hyperparameter. We therefore use $0.50$ as the default retention ratio, which
offers the best retrieval performance while maintaining strong classification
and localization results with moderate teacher side computation.
\section{Conclusion}

We presented an attribution guided shared-private decoupling framework for noise reduced audio-visual representation learning. Our key idea is to identify shared-cover subviews with layer-truncated attribution, and to use the resulting high-relevance and low-relevance teacher responses to supervise decoupled shared and private student tokens. This allows the final model to internalize fine-grained attribution guidance without introducing extra inference cost. Experiments on cross-modal retrieval, audio-visual classification, and sound-prompted semantic segmentation demonstrate that our method improves both global audio-visual alignment and patch-level semantic correspondence. We hope that this work encourages further exploration of token-level supervision and shared-private decomposition in multimodal representation learning.

\begin{ack}
This work was supported by National Key R\&D Program of China under Grant No.~2023ZD0120400, in part by National Natural Science Foundation of China (No.~62472423, 62276260).
\end{ack}

\clearpage
{\small
\bibliographystyle{unsrt}   %
\bibliography{main}         %

@String(CVPR= {IEEE Conf. Comput. Vis. Pattern Recog.})

@String(ICCV= {Int. Conf. Comput. Vis.})

@String(ECCV= {Eur. Conf. Comput. Vis.})

@String(NIPS= {Adv. Neural Inform. Process. Syst.})

@String(ICASSP=	{ICASSP})

@String(AAAI = {AAAI})

@String(CVPR  = {CVPR})

@String(ICCV  = {ICCV})

@String(ECCV  = {ECCV})

@String(NIPS  = {NeurIPS})

@inproceedings{gong2022cavmae,
    title={Contrastive Audio-Visual Masked Autoencoder},
    author={Yuan Gong and Andrew Rouditchenko and Alexander H. Liu and David Harwath and Leonid Karlinsky and Hilde Kuehne and James R. Glass},
    booktitle={The Eleventh International Conference on Learning Representations},
    year={2023},
    url={https://openreview.net/forum?id=QPtMRyk5rb}
}

@inproceedings{araujo2025cav,
  title={CAV-MAE Sync: Improving Contrastive Audio-Visual Mask Autoencoders via Fine-Grained Alignment},
  author={Araujo, Edson and Rouditchenko, Andrew and Gong, Yuan and Bhati, Saurabhchand and Thomas, Samuel and Kingsbury, Brian and Karlinsky, Leonid and Feris, Rogerio and Glass, James R and Kuehne, Hilde},
  booktitle={Proceedings of the Computer Vision and Pattern Recognition Conference},
  pages={18794--18803},
  year={2025}
}

@inproceedings{Dufumier2025PID,
 author = {Dufumier, Benoit and Castillo Navarro, Javiera and Tuia, Devis and Thiran, Jean-Philippe},
 booktitle = {International Conference on Learning Representations},
 editor = {Y. Yue and A. Garg and N. Peng and F. Sha and R. Yu},
 pages = {5408--5432},
 title = {What to align in multimodal contrastive learning?},
 url = {https://proceedings.iclr.cc/paper_files/paper/2025/file/108030643e640ac050e0ed5e6aace48f-Paper-Conference.pdf},
 volume = {2025},
 year = {2025}
}

@INPROCEEDINGS{DeepIB,
  author={Tishby, Naftali and Zaslavsky, Noga},
  booktitle={2015 IEEE Information Theory Workshop (ITW)}, 
  title={Deep learning and the information bottleneck principle}, 
  year={2015},
  volume={},
  number={},
  pages={1-5},
  doi={10.1109/ITW.2015.7133169}}

@article{Bousselham2024LeGrad,
  title={LeGrad: An Explainability Method for Vision Transformers via Feature Formation Sensitivity},
  author={Walid Bousselham and Angie Boggust and Sofian Chaybouti and Hendrik Strobelt and Hilde Kuehne},
  journal={ArXiv},
  year={2024},
  volume={abs/2404.03214},
  url={https://api.semanticscholar.org/CorpusID:268889646}
}

@InProceedings{achtibat24AttnLRP,
  title = 	 {{A}ttn{LRP}: Attention-Aware Layer-Wise Relevance Propagation for Transformers},
  author =       {Achtibat, Reduan and Hatefi, Sayed Mohammad Vakilzadeh and Dreyer, Maximilian and Jain, Aakriti and Wiegand, Thomas and Lapuschkin, Sebastian and Samek, Wojciech},
  booktitle = 	 {Proceedings of the 41st International Conference on Machine Learning},
  pages = 	 {135--168},
  year = 	 {2024},
  editor = 	 {Salakhutdinov, Ruslan and Kolter, Zico and Heller, Katherine and Weller, Adrian and Oliver, Nuria and Scarlett, Jonathan and Berkenkamp, Felix},
  volume = 	 {235},
  series = 	 {Proceedings of Machine Learning Research},
  month = 	 {21--27 Jul},
  publisher =    {PMLR},
  url = 	 {https://proceedings.mlr.press/v235/achtibat24a.html}
}

@inproceedings{chefer2021transformer,
  title={Transformer interpretability beyond attention visualization},
  author={Chefer, Hila and Gur, Shir and Wolf, Lior},
  booktitle={Proceedings of the IEEE/CVF conference on computer vision and pattern recognition},
  pages={782--791},
  year={2021}
}

@article{Chefer2021GenericAE,
  title={Generic Attention-model Explainability for Interpreting Bi-Modal and Encoder-Decoder Transformers},
  author={Hila Chefer and Shir Gur and Lior Wolf},
  journal={2021 IEEE/CVF International Conference on Computer Vision (ICCV)},
  year={2021},
  pages={387-396},
  url={https://api.semanticscholar.org/CorpusID:232417173}
}

@misc{linge26,
      title={Semantic Noise Reduction via Teacher-Guided Dual-Path Audio-Visual Representation Learning}, 
      author={Linge Wang and Yingying Chen and Bingke Zhu and Lu Zhou and Jinqiao Wang},
      year={2026},
      eprint={2604.08147},
      archivePrefix={arXiv},
      primaryClass={cs.SD},
      url={https://arxiv.org/abs/2604.08147}, 
}

@article{huang2023mavil,
  title={Mavil: Masked audio-video learners},
  author={Huang, Po-Yao and Sharma, Vasu and Xu, Hu and Ryali, Chaitanya and Li, Yanghao and Li, Shang-Wen and Ghosh, Gargi and Malik, Jitendra and Feichtenhofer, Christoph and others},
  journal={Advances in Neural Information Processing Systems},
  volume={36},
  pages={20371--20393},
  year={2023}
}

@inproceedings{girdhar2023imagebind,
  title={Imagebind: One embedding space to bind them all},
  author={Girdhar, Rohit and El-Nouby, Alaaeldin and Liu, Zhuang and Singh, Mannat and Alwala, Kalyan Vasudev and Joulin, Armand and Misra, Ishan},
  booktitle={Proceedings of the IEEE/CVF conference on computer vision and pattern recognition},
  pages={15180--15190},
  year={2023}
}

@inproceedings{georgescu2023audiovisual,
  title={Audiovisual masked autoencoders},
  author={Georgescu, Mariana-Iuliana and Fonseca, Eduardo and Ionescu, Radu Tudor and Lucic, Mario and Schmid, Cordelia and Arnab, Anurag},
  booktitle={Proceedings of the IEEE/CVF International Conference on Computer Vision},
  pages={16144--16154},
  year={2023}
}

@inproceedings{he2022masked,
  title={Masked autoencoders are scalable vision learners},
  author={He, Kaiming and Chen, Xinlei and Xie, Saining and Li, Yanghao and Doll{\'a}r, Piotr and Girshick, Ross},
  booktitle={Proceedings of the IEEE/CVF conference on computer vision and pattern recognition},
  pages={16000--16009},
  year={2022}
}

@article{oord2018representation,
  title={Representation learning with contrastive predictive coding},
  author={Oord, Aaron van den and Li, Yazhe and Vinyals, Oriol},
  journal={arXiv preprint arXiv:1807.03748},
  year={2018}
}

@misc{tishby2000ori_IB,
      title={The information bottleneck method}, 
      author={Naftali Tishby and Fernando C. Pereira and William Bialek},
      year={2000},
      eprint={physics/0004057},
      archivePrefix={arXiv},
      primaryClass={physics.data-an},
      url={https://arxiv.org/abs/physics/0004057}, 
}

@article{aytar2016soundnet,
  title={Soundnet: Learning sound representations from unlabeled video},
  author={Aytar, Yusuf and Vondrick, Carl and Torralba, Antonio},
  journal={Advances in neural information processing systems},
  volume={29},
  year={2016}
}

@inproceedings{guo2024crossmae,
  title={Crossmae: Cross-modality masked autoencoders for region-aware audio-visual pre-training},
  author={Guo, Yuxin and Sun, Siyang and Ma, Shuailei and Zheng, Kecheng and Bao, Xiaoyi and Ma, Shijie and Zou, Wei and Zheng, Yun},
  booktitle={Proceedings of the IEEE/CVF Conference on Computer Vision and Pattern Recognition},
  pages={26721--26731},
  year={2024}
}

@inproceedings{ngiam2011multimodal,
author = {Ngiam, Jiquan and Khosla, Aditya and Kim, Mingyu and Nam, Juhan and Lee, Honglak and Ng, Andrew Y.},
title = {Multimodal deep learning},
year = {2011},
isbn = {9781450306195},
publisher = {Omnipress},
address = {Madison, WI, USA},
booktitle = {Proceedings of the 28th International Conference on International Conference on Machine Learning},
pages = {689–696},
numpages = {8},
location = {Bellevue, Washington, USA},
series = {ICML'11}
}

@article{srivastava2012multimodal,
  title={Multimodal learning with deep boltzmann machines},
  author={Srivastava, Nitish and Salakhutdinov, Russ R},
  journal={Advances in neural information processing systems},
  volume={25},
  year={2012}
}

@article{frome2013devise,
  title={Devise: A deep visual-semantic embedding model},
  author={Frome, Andrea and Corrado, Greg S and Shlens, Jon and Bengio, Samy and Dean, Jeff and Ranzato, Marc'Aurelio and Mikolov, Tomas},
  journal={Advances in neural information processing systems},
  volume={26},
  year={2013}
}

@article{kiros2014unifying,
  title={Unifying visual-semantic embeddings with multimodal neural language models},
  author={Kiros, Ryan and Salakhutdinov, Ruslan and Zemel, Richard S},
  journal={arXiv preprint arXiv:1411.2539},
  year={2014}
}

@article{lu2019vilbert,
  title={Vilbert: Pretraining task-agnostic visiolinguistic representations for vision-and-language tasks},
  author={Lu, Jiasen and Batra, Dhruv and Parikh, Devi and Lee, Stefan},
  journal={Advances in neural information processing systems},
  volume={32},
  year={2019}
}

@inproceedings{radford2021learning,
  title={Learning transferable visual models from natural language supervision},
  author={Radford, Alec and Kim, Jong Wook and Hallacy, Chris and Ramesh, Aditya and Goh, Gabriel and Agarwal, Sandhini and Sastry, Girish and Askell, Amanda and Mishkin, Pamela and Clark, Jack and others},
  booktitle={International conference on machine learning},
  pages={8748--8763},
  year={2021},
  organization={PmLR}
}

@article{williams2010nonnegative,
  title={Nonnegative decomposition of multivariate information},
  author={Williams, Paul L and Beer, Randall D},
  journal={arXiv preprint arXiv:1004.2515},
  year={2010}
}

@article{bertschinger2014quantifying,
  title={Quantifying unique information},
  author={Bertschinger, Nils and Rauh, Johannes and Olbrich, Eckehard and Jost, J{\"u}rgen and Ay, Nihat},
  journal={Entropy},
  volume={16},
  number={4},
  pages={2161--2183},
  year={2014},
  publisher={Molecular Diversity Preservation International (MDPI)}
}

@inproceedings{gao2025disentangled,
  title={Disentangled Cross-Modal Representation Learning with Enhanced Mutual Supervision},
  author={Gao, Lu and Chen, Wenlan and Wang, Daoyuan and Guo, Fei and Liang, Cheng},
  booktitle={The Thirty-ninth Annual Conference on Neural Information Processing Systems},
  year={2025}
}

@inproceedings{liuplug,
  title={Plug-and-play Feature Causality Decomposition for Multimodal Representation Learning},
  author={Liu, Ye and Ji, Zihan and Cai, Hongmin},
  booktitle={The Thirty-ninth Annual Conference on Neural Information Processing Systems},
  year={2025}
}

@article{xia2023achieving,
  title={Achieving cross modal generalization with multimodal unified representation},
  author={Xia, Yan and Huang, Hai and Zhu, Jieming and Zhao, Zhou},
  journal={Advances in Neural Information Processing Systems},
  volume={36},
  pages={63529--63541},
  year={2023}
}

@inproceedings{gemmeke2017audio,
  title={Audio set: An ontology and human-labeled dataset for audio events},
  author={Gemmeke, Jort F and Ellis, Daniel PW and Freedman, Dylan and Jansen, Aren and Lawrence, Wade and Moore, R Channing and Plakal, Manoj and Ritter, Marvin},
  booktitle={2017 IEEE international conference on acoustics, speech and signal processing (ICASSP)},
  pages={776--780},
  year={2017},
  organization={IEEE}
}

@inproceedings{chen2020vggsound,
  title={Vggsound: A large-scale audio-visual dataset},
  author={Chen, Honglie and Xie, Weidi and Vedaldi, Andrea and Zisserman, Andrew},
  booktitle={ICASSP 2020-2020 IEEE International Conference on Acoustics, Speech and Signal Processing (ICASSP)},
  pages={721--725},
  year={2020},
  organization={IEEE}
}

@inproceedings{AvsBench,
author = {Zhou, Jinxing and Wang, Jianyuan and Zhang, Jiayi and Sun, Weixuan and Zhang, Jing and Birchfield, Stan and Guo, Dan and Kong, Lingpeng and Wang, Meng and Zhong, Yiran},
title = {Audio–Visual Segmentation},
year = {2022},
isbn = {978-3-031-19835-9},
publisher = {Springer-Verlag},
address = {Berlin, Heidelberg},
url = {https://doi.org/10.1007/978-3-031-19836-6_22},
doi = {10.1007/978-3-031-19836-6_22},
booktitle = {Computer Vision – ECCV 2022: 17th European Conference, Tel Aviv, Israel, October 23–27, 2022, Proceedings, Part XXXVII},
pages = {386–403},
numpages = {18},
location = {Tel Aviv, Israel}
}

@article{zhu2024languagebind,
  title={Languagebind: Extending video-language pretraining to n-modality by language-based semantic alignment},
  author={Zhu, Bin and Lin, Bin and Ning, Munan and Yan, Yang and Cui, Jiaxi and Wang, HongFa and Pang, Yatian and Jiang, Wenhao and Zhang, Junwu and Li, Zongwei and others},
  journal={arXiv preprint arXiv:2310.01852},
  year={2023}
}

@inproceedings{lin2024avsiam,
  title={Siamese vision transformers are scalable audio-visual learners},
  author={Lin, Yan-Bo and Bertasius, Gedas},
  booktitle={European Conference on Computer Vision},
  pages={303--321},
  year={2024},
  organization={Springer}
}

@inproceedings{hamilton2024denseav,
  title={Separating the" chirp" from the" chat": Self-supervised visual grounding of sound and language},
  author={Hamilton, Mark and Zisserman, Andrew and Hershey, John R and Freeman, William T},
  booktitle={Proceedings of the IEEE/CVF conference on computer vision and pattern recognition},
  pages={13117--13127},
  year={2024}
}

@inproceedings{Probabalistic,
author = {Fisher, John W. and Darrell, Trevor},
title = {Probabalistic Models and Informative Subspaces for Audiovisual Correspondence},
year = {2002},
isbn = {3540437460},
publisher = {Springer-Verlag},
address = {Berlin, Heidelberg},
booktitle = {Proceedings of the 7th European Conference on Computer Vision-Part III},
pages = {592–603},
numpages = {12},
series = {ECCV '02}
}

@inproceedings{statistical,
author = {Fisher, John W. and Darrell, Trevor and Freeman, William T. and Viola, Paul},
title = {Learning joint statistical models for audio-visual fusion and segregation},
year = {2000},
publisher = {MIT Press},
address = {Cambridge, MA, USA},
booktitle = {Proceedings of the 14th International Conference on Neural Information Processing Systems},
pages = {742–748},
numpages = {7},
location = {Denver, CO},
series = {NIPS'00}
}

@inproceedings{arandjelovic2018objects,
  title={Objects that sound},
  author={Arandjelovic, Relja and Zisserman, Andrew},
  booktitle={Proceedings of the European conference on computer vision (ECCV)},
  pages={435--451},
  year={2018}
}

@misc{DAVENET,
      title={Jointly Discovering Visual Objects and Spoken Words from Raw Sensory Input}, 
      author={David Harwath and Adrià Recasens and Dídac Surís and Galen Chuang and Antonio Torralba and James Glass},
      year={2018},
      eprint={1804.01452},
      archivePrefix={arXiv},
      primaryClass={cs.CV},
      url={https://arxiv.org/abs/1804.01452}, 
}

@misc{owens2016ambientsoundprovidessupervision,
      title={Ambient Sound Provides Supervision for Visual Learning}, 
      author={Andrew Owens and Jiajun Wu and Josh H. McDermott and William T. Freeman and Antonio Torralba},
      year={2016},
      eprint={1608.07017},
      archivePrefix={arXiv},
      primaryClass={cs.CV},
      url={https://arxiv.org/abs/1608.07017}, 
}

@inproceedings{arandjelovic2017look,
  title={Look, listen and learn},
  author={Arandjelovic, Relja and Zisserman, Andrew},
  booktitle={Proceedings of the IEEE international conference on computer vision},
  pages={609--617},
  year={2017}
}

@inproceedings{Lee2021CrossAttentionalAF,
  title={Cross-Attentional Audio-Visual Fusion for Weakly-Supervised Action Localization},
  author={Jun-Tae Lee and Mihir Jain and Hyoungwoo Park and Sungrack Yun},
  booktitle={International Conference on Learning Representations},
  year={2021},
  url={https://api.semanticscholar.org/CorpusID:235614339}
}

@inproceedings{morgado2021crossmodal,
  title={Audio-visual instance discrimination with cross-modal agreement},
  author={Morgado, Pedro and Vasconcelos, Nuno and Misra, Ishan},
  booktitle={Proceedings of the IEEE/CVF conference on computer vision and pattern recognition},
  pages={12475--12486},
  year={2021}
}

@article{DTD,
author = {Montavon, Gr\'{e}goire and Lapuschkin, Sebastian and Binder, Alexander and Samek, Wojciech and M\"{u}ller, Klaus-Robert},
title = {Explaining nonlinear classification decisions with deep Taylor decomposition},
year = {2017},
issue_date = {May 2017},
publisher = {Elsevier Science Inc.},
address = {USA},
volume = {65},
number = {C},
issn = {0031-3203},
url = {https://doi.org/10.1016/j.patcog.2016.11.008},
doi = {10.1016/j.patcog.2016.11.008},
journal = {Pattern Recogn.},
month = may,
pages = {211–222},
numpages = {12}
}

@inproceedings{abnar-zuidema-2020-quantifying,
    title = "Quantifying Attention Flow in Transformers",
    author = "Abnar, Samira  and
      Zuidema, Willem",
    editor = "Jurafsky, Dan  and
      Chai, Joyce  and
      Schluter, Natalie  and
      Tetreault, Joel",
    booktitle = "Proceedings of the 58th Annual Meeting of the Association for Computational Linguistics",
    month = jul,
    year = "2020",
    address = "Online",
    publisher = "Association for Computational Linguistics",
    url = "https://aclanthology.org/2020.acl-main.385/",
    doi = "10.18653/v1/2020.acl-main.385",
    pages = "4190--4197"
}

@book{molnar2020interpretable,
  title={Interpretable machine learning},
  author={Molnar, Christoph},
  year={2020},
  publisher={Lulu. com}
}

@inproceedings{ribeiro2016should,
  title={" Why should i trust you?" Explaining the predictions of any classifier},
  author={Ribeiro, Marco Tulio and Singh, Sameer and Guestrin, Carlos},
  booktitle={Proceedings of the 22nd ACM SIGKDD international conference on knowledge discovery and data mining},
  pages={1135--1144},
  year={2016}
}

@article{erhan2009visualizing,
  title={Visualizing higher-layer features of a deep network},
  author={Erhan, Dumitru and Bengio, Yoshua and Courville, Aaron and Vincent, Pascal},
  journal={University of Montreal},
  volume={1341},
  number={3},
  pages={1},
  year={2009}
}

@inproceedings{sundararajan2017axiomatic,
  title={Axiomatic attribution for deep networks},
  author={Sundararajan, Mukund and Taly, Ankur and Yan, Qiqi},
  booktitle={International conference on machine learning},
  pages={3319--3328},
  year={2017},
  organization={PMLR}
}

@inproceedings{zeiler2014visualizing,
  title={Visualizing and understanding convolutional networks},
  author={Zeiler, Matthew D and Fergus, Rob},
  booktitle={European conference on computer vision},
  pages={818--833},
  year={2014},
  organization={Springer}
}

@inproceedings{selvaraju2017grad,
  title={Grad-cam: Visual explanations from deep networks via gradient-based localization},
  author={Selvaraju, Ramprasaath R and Cogswell, Michael and Das, Abhishek and Vedantam, Ramakrishna and Parikh, Devi and Batra, Dhruv},
  booktitle={Proceedings of the IEEE international conference on computer vision},
  pages={618--626},
  year={2017}
}

@inproceedings{jain-wallace-2019-attention,
    title = "{A}ttention is not {E}xplanation",
    author = "Jain, Sarthak  and
      Wallace, Byron C.",
    editor = "Burstein, Jill  and
      Doran, Christy  and
      Solorio, Thamar",
    booktitle = "Proceedings of the 2019 Conference of the North {A}merican Chapter of the Association for Computational Linguistics: Human Language Technologies, Volume 1 (Long and Short Papers)",
    month = jun,
    year = "2019",
    address = "Minneapolis, Minnesota",
    publisher = "Association for Computational Linguistics",
    url = "https://aclanthology.org/N19-1357/",
    doi = "10.18653/v1/N19-1357",
    pages = "3543--3556"
}

@inproceedings{serrano-smith-2019-attention,
    title = "Is Attention Interpretable?",
    author = "Serrano, Sofia  and
      Smith, Noah A.",
    editor = "Korhonen, Anna  and
      Traum, David  and
      M{\`a}rquez, Llu{\'i}s",
    booktitle = "Proceedings of the 57th Annual Meeting of the Association for Computational Linguistics",
    month = jul,
    year = "2019",
    address = "Florence, Italy",
    publisher = "Association for Computational Linguistics",
    url = "https://aclanthology.org/P19-1282/",
    doi = "10.18653/v1/P19-1282",
    pages = "2931--2951"
}

@article{Hao_Dong_Wei_Xu_2021, title={Self-Attention Attribution: Interpreting Information Interactions Inside Transformer}, volume={35}, url={https://ojs.aaai.org/index.php/AAAI/article/view/17533}, DOI={10.1609/aaai.v35i14.17533}, abstractNote={The great success of Transformer-based models benefits from the powerful multi-head self-attention mechanism, which learns token dependencies and encodes contextual information from the input. Prior work strives to attribute model decisions to individual input features with different saliency measures, but they fail to explain how these input features interact with each other to reach predictions. In this paper, we propose a self-attention attribution method to interpret the information interactions inside Transformer. We take BERT as an example to conduct extensive studies. Firstly, we apply self-attention attribution to identify the important attention heads, while others can be pruned with marginal performance degradation. Furthermore, we extract the most salient dependencies in each layer to construct an attribution tree, which reveals the hierarchical interactions inside Transformer. Finally, we show that the attribution results can be used as adversarial patterns to implement non-targeted attacks towards BERT.}, number={14}, journal={Proceedings of the AAAI Conference on Artificial Intelligence}, author={Hao, Yaru and Dong, Li and Wei, Furu and Xu, Ke}, year={2021}, month={May}, pages={12963-12971} }

@article{
chen2023beyond,
title={Beyond Intuition: Rethinking Token Attributions inside Transformers},
author={Jiamin Chen and Xuhong Li and Lei Yu and Dejing Dou and Haoyi Xiong},
journal={Transactions on Machine Learning Research},
issn={2835-8856},
year={2023},
url={https://openreview.net/forum?id=rm0zIzlhcX},
note={}
}

@inproceedings{10.5555/3327757.3327874,
author = {Korbar, Bruno and Tran, Du and Torresani, Lorenzo},
title = {Cooperative learning of audio and video models from self-supervised synchronization},
year = {2018},
publisher = {Curran Associates Inc.},
address = {Red Hook, NY, USA},
booktitle = {Proceedings of the 32nd International Conference on Neural Information Processing Systems},
pages = {7774–7785},
numpages = {12},
location = {Montr\'{e}al, Canada},
series = {NIPS'18}
}

@inproceedings{NEURIPS2020_6f2268bd,
 author = {Alwassel, Humam and Mahajan, Dhruv and Korbar, Bruno and Torresani, Lorenzo and Ghanem, Bernard and Tran, Du},
 booktitle = {Advances in Neural Information Processing Systems},
 editor = {H. Larochelle and M. Ranzato and R. Hadsell and M.F. Balcan and H. Lin},
 pages = {9758--9770},
 publisher = {Curran Associates, Inc.},
 title = {Self-Supervised Learning by Cross-Modal Audio-Video Clustering},
 url = {https://proceedings.neurips.cc/paper_files/paper/2020/file/6f2268bd1d3d3ebaabb04d6b5d099425-Paper.pdf},
 volume = {33},
 year = {2020}
}

@INPROCEEDINGS{ADE20k,
  author={Zhou, Bolei and Zhao, Hang and Puig, Xavier and Fidler, Sanja and Barriuso, Adela and Torralba, Antonio},
  booktitle={2017 IEEE Conference on Computer Vision and Pattern Recognition (CVPR)}, 
  title={Scene Parsing through ADE20K Dataset}, 
  year={2017},
  volume={},
  number={},
  pages={5122-5130},
  doi={10.1109/CVPR.2017.544}}
}

\newpage
\appendix
\section{Proof of Proposition~\ref{prop:shared_cover_ib}}
\label{app:proof_subview_ib}

We provide the proof for the audio modality, while the same argument applies to the visual modality.
Recall that the full audio input is decomposed into the selected shared-cover subview and its discarded complement:
\begin{equation}
    A = A_{\mathcal{T}} \cup A_{\mathcal{R}}, 
    \qquad 
    A_{\mathcal{T}} \cap A_{\mathcal{R}} = \emptyset .
\end{equation}
For any token subset $B \subseteq A$, the input-level IB criterion is defined as
\begin{equation}
    F(B) = I(B;S) - \beta I(B;P_A),
\end{equation}
where $S$ denotes the cross-modal shared information and $P_A$ denotes the audio-private residual information.

Using the chain rule of mutual information, we have
\begin{equation}
    I(A;S)
    =
    I(A_{\mathcal{T}};S)
    +
    I(A_{\mathcal{R}};S \mid A_{\mathcal{T}}),
\end{equation}
and similarly,
\begin{equation}
    I(A;P_A)
    =
    I(A_{\mathcal{T}};P_A)
    +
    I(A_{\mathcal{R}};P_A \mid A_{\mathcal{T}}).
\end{equation}
Substituting the two decompositions into $F(A)$ gives
\begin{align}
    F(A)
    &=
    I(A;S) - \beta I(A;P_A) \nonumber \\
    &=
    I(A_{\mathcal{T}};S)
    +
    I(A_{\mathcal{R}};S \mid A_{\mathcal{T}})
    -
    \beta I(A_{\mathcal{T}};P_A)
    -
    \beta I(A_{\mathcal{R}};P_A \mid A_{\mathcal{T}}) \nonumber \\
    &=
    F(A_{\mathcal{T}})
    +
    I(A_{\mathcal{R}};S \mid A_{\mathcal{T}})
    -
    \beta I(A_{\mathcal{R}};P_A \mid A_{\mathcal{T}}).
\end{align}
Therefore,
\begin{equation}
    F(A_{\mathcal{T}}) - F(A)
    =
    \beta I(A_{\mathcal{R}};P_A \mid A_{\mathcal{T}})
    -
    I(A_{\mathcal{R}};S \mid A_{\mathcal{T}}).
\end{equation}

By the Shared Cover Condition,
\begin{equation}
    I(A_{\mathcal{R}};S \mid A_{\mathcal{T}}) \leq \epsilon_S,
\end{equation}
and by the Private Tail Condition,
\begin{equation}
    I(A_{\mathcal{R}};P_A \mid A_{\mathcal{T}}) \geq \epsilon_P.
\end{equation}
It follows that
\begin{equation}
    F(A_{\mathcal{T}}) - F(A)
    \geq
    \beta \epsilon_P - \epsilon_S.
\end{equation}
Consequently, if $\beta \epsilon_P > \epsilon_S$, then
\begin{equation}
    F(A_{\mathcal{T}}) > F(A).
\end{equation}
This proves Proposition~\ref{prop:shared_cover_ib}.

The equality above also reveals the exact condition under which a selected subview improves the input-level IB objective:
\begin{equation}
    F(A_{\mathcal{T}}) > F(A)
    \quad \Longleftrightarrow \quad
    \beta I(A_{\mathcal{R}};P_A \mid A_{\mathcal{T}})
    >
    I(A_{\mathcal{R}};S \mid A_{\mathcal{T}}).
\end{equation}
In other words, the selected subview is more suitable than the full input for learning the shared representation when the discarded tokens contain sufficiently more conditional private residual information than conditional shared information, after weighting by $\beta$.
\section{Implementation Details}
\label{app:implementation}
\subsection{Datasets Details}
\paragraph{AudioSet.}
AudioSet-2M~\cite{gemmeke2017audio} is a large-scale audio-visual dataset originally curated for audio event classification. 
It contains approximately two million 10-second video clips collected from YouTube, with audio event annotations over 527 classes. 
Due to its large scale and diverse acoustic events, AudioSet-2M has been widely used for audio-visual self-supervised pretraining. 
However, the dataset was not explicitly designed for audio-visual correspondence learning. 
Therefore, some clips contain weakly correlated or mismatched audio-visual content, such as off-screen sound sources, silent visual regions, or background audio unrelated to the visible scene. 
This property makes AudioSet-2M a realistic but noisy pretraining source for cross-modal representation learning.

In practice, not all AudioSet videos are accessible because the original data are hosted on YouTube and may become unavailable due to removal, privacy settings, or regional restrictions. 
In our implementation, approximately 350K samples could not be downloaded. 
Among the downloaded videos, we further remove clips with missing audio tracks, corrupted frames, or invalid decoding results. 
After filtering, we obtain 1,390,395 valid audio-visual pairs for pretraining. 
For downstream audio-visual classification, we use AudioSet-20K following the standard probing protocol. 
For zero-shot cross-modal retrieval, we follow the AudioSet evaluation split provided by CAV-MAE~\cite{gong2022cavmae}.

\paragraph{VGGSound.}
VGGSound~\cite{chen2020vggsound} is a large-scale audio-visual dataset collected from YouTube, containing around 200K 10-second video clips across 309 sound event categories. 
Unlike AudioSet, VGGSound was curated with explicit audio-visual correspondence in mind: each clip is expected to contain a visible object or event that is semantically related to the accompanying sound. 
This property makes VGGSound a relatively clean benchmark for evaluating audio-visual alignment, since it contains fewer off-screen or weakly matched audio-visual pairs. 
In our experiments, we use VGGSound for both downstream classification and zero-shot cross-modal retrieval evaluation. 
Specifically, we train the probing classifier on the VGGSound training split, and follow the retrieval evaluation split provided by CAV-MAE Sync~\cite{araujo2025cav} for zero-shot audio-visual retrieval.

\paragraph{ADE20k-Sound Prompted}
ADE20K-Sound is a sound-prompted semantic segmentation benchmark constructed from ADE20K~\cite{ADE20k} and VGGSound~\cite{chen2020vggsound}. 
It contains 106 images covering 20 ADE20K object categories, where each image is paired with a semantically corresponding sound clip sampled from VGGSound. 
The benchmark is designed to evaluate whether an audio-visual model can localize image regions that are relevant to the given sound prompt. 

\paragraph{AVSBench}
AVSBench~\cite{AvsBench} is a pixel-level audio-visual segmentation benchmark designed to evaluate whether a model can localize sounding objects in videos. 
Different from ADE20K-Sound, which is constructed from static images paired with sound prompts, AVSBench provides video-based audio-visual samples with manually annotated segmentation masks for sound-producing objects. 
The original AVSBench-object benchmark contains two settings: single sound source segmentation (S4) and multiple sound source segmentation (MS3), covering 5,356 videos and 23 object categories in total. 
This makes it a larger and more dedicated benchmark for evaluating fine-grained audio-visual localization. 

\subsection{Training Pipeline and Implementation Details}
\label{app:training_details}

\paragraph{Preprocessing.}
We follow the CAV-MAE Sync preprocessing pipeline for both pretraining and downstream evaluation. 
Given a video, we uniformly sample $T=16$ frames. 
Each frame is resized and center-cropped to $224\times224$, and then patchified into $16\times16$ visual patches, producing $196$ visual patch tokens per frame. 
The audio waveform is converted into 128-bin log-Mel filterbanks using a 25 ms Hanning window and a 10 ms hop size. 
For each sampled frame, we extract a temporally aligned 4-second spectrogram segment of size $128\times416$ by mapping the frame timestamp to the corresponding spectrogram coordinate. 
Each spectrogram segment is also patchified with a $16\times16$ patch size, resulting in $208$ audio patch tokens. 
Standard normalization is applied to the audio spectrograms with mean $-5.081$ and standard deviation $4.485$.

During pretraining, we randomly sample one frame and its temporally aligned spectrogram segment from the $T=16$ candidates as the input pair. 
During downstream classification, all $T=16$ frame-spectrogram pairs are used to construct a temporal sequence for attention probing.

\paragraph{Dual-path pretraining backbone.}
Our training framework inherits the dual-path design of TG-DP~\cite{linge26}, which separates masked reconstruction and contrastive alignment into two objective-specific forward passes. 
For each audio-visual pair, the model processes the input twice with different masking configurations. 
The reconstruction branch adopts a high masking ratio of 0.75 and optimizes masked autoencoding, while the contrastive branch adopts a lower masking ratio of 0.50 and optimizes cross-modal alignment. 
The two branches share the same modality-specific encoders and unified block, but their losses are computed from different masked views.

In the reconstruction branch, audio and visual patch tokens are randomly masked and then encoded by the modality-specific encoders. 
The visible audio and visual tokens are fed into the unified block and decoded to reconstruct the masked patches. 
The reconstruction loss is defined as
\begin{equation}
    \mathcal{L}_{\mathrm{rec}}
    =
    \sum_{m\in\{a,v\}}
    \frac{1}{|\mathcal{M}^{m}|}
    \sum_{i\in \mathcal{M}^{m}}
    \left\|
    \hat{x}^{m}_{i} - x^{m}_{i}
    \right\|_{2}^{2},
\end{equation}
where $\mathcal{M}^{m}$ denotes the masked token set of modality $m$, and $\hat{x}^{m}_{i}$ is the reconstructed patch token.

In the contrastive branch, the student receives a lower-mask-ratio view and produces modality-level representations for cross-modal alignment. 
Different from the original single-global-token design, our student uses two learnable tokens for each modality: a shared token $z_s^m$ and a private token $z_p^m$. 
The shared tokens are used for cross-modal contrastive learning:
\begin{equation}
    \mathcal{L}_{\mathrm{cl}}
    =
    \frac{1}{2}
    \left(
    \mathcal{L}_{a\rightarrow v}
    +
    \mathcal{L}_{v\rightarrow a}
    \right),
\end{equation}
where
\begin{equation}
    \mathcal{L}_{a\rightarrow v}
    =
    -
    \frac{1}{N}
    \sum_{i=1}^{N}
    \log
    \frac{
    \exp(\langle z^{a}_{s,i}, z^{v}_{s,i}\rangle / \tau)
    }{
    \sum_{j=1}^{N}
    \exp(\langle z^{a}_{s,i}, z^{v}_{s,j}\rangle / \tau)
    },
\end{equation}
and $\mathcal{L}_{v\rightarrow a}$ is defined symmetrically. 
Here, $N$ is the batch size and $\tau$ is the temperature parameter. 
The base dual-path objective is therefore written as
\begin{equation}
    \mathcal{L}_{\mathrm{base}}
    =
    \lambda_{\mathrm{rec}}\mathcal{L}_{\mathrm{rec}}
    +
    \lambda_{\mathrm{cl}}\mathcal{L}_{\mathrm{cl}}.
\end{equation}

\paragraph{Early-converged relevance teacher.}
Before student training, we pretrain a relevance teacher using the same dual-path backbone as TG-DP~\cite{linge26}. 
The teacher is trained from scratch with the original random masking strategy, without LTA-based subview construction or shared-private token decoupling. 
The full TG-DP training schedule contains 35 epochs, while we stop the teacher at epoch 18. 
This early stopping is motivated by the empirical observation that token-relevance rankings become sufficiently stable in the middle stage of training, so a fully converged teacher is not necessary for generating subview supervision. 
After pretraining, the teacher is frozen and used only to compute LTA scores and produce stop-gradient anchor targets during student training.

\paragraph{Attribution-guided student training.}
The student is initialized from scratch and trained for 25 epochs using the same dual-path backbone. 
For each audio-visual pair, the frozen teacher computes LTA scores from the last unified block and ranks patch tokens within each modality. 
We use a retention ratio of $\rho=0.50$ to split tokens into a high-relevance shared-cover subview and a low-relevance residual complement. 
The teacher responses on these two subviews serve as stop-gradient anchors for the student's shared and private tokens, respectively.

\paragraph{Downstream evaluation.}
For zero-shot cross-modal retrieval, we use the shared token $z_s^m$ as the global representation of each modality, since retrieval mainly evaluates cross-modal shared correspondence. 
For audio-visual classification, we follow the attention probing protocol and freeze the pretrained encoders. 
At each of the $T=16$ temporal steps, we average the shared and private tokens to form a single modality-level token, concatenate the audio and visual tokens over time, and train a lightweight attention-based classification head. 
For sound-prompted semantic segmentation, we use the similarity between the sound representation and visual patch tokens to produce the localization heatmap, following the zero-shot evaluation protocol used in prior audio-visual grounding benchmarks.

Table~\ref{tab:pretrain-lp} summarizes the main hyperparameters used for teacher pretraining, student pretraining, and downstream probing.

\begin{table}[t]
\centering
\small
\setlength{\tabcolsep}{3.5pt}
\renewcommand{\arraystretch}{1.05}
\resizebox{\linewidth}{!}{
\begin{tabular}{lcccc}
\toprule
\textbf{Setting} 
& \textbf{Teacher Pretrain} 
& \textbf{Student Pretrain} 
& \shortstack{\textbf{AS20K}\\\textbf{Attn. Probe}} 
& \shortstack{\textbf{VGGSound}\\\textbf{Attn. Probe}} \\
\midrule
Dataset & AS-2M & AS-2M & AS-20K & VGGSound \\
Optimizer & Adam & Adam & Adam & Adam \\
Learning Rate & $2\!\times\!10^{-4}$ & $2\!\times\!10^{-4}$ & $5\!\times\!10^{-3}$ & $1\!\times\!10^{-3}$ \\
LR Scheduler & Cosine & Cosine & Cosine & Cosine \\
Epochs & 18 & 25 & 10 & 10 \\
Warmup Epochs & 3.5 & 2.5 & 1.5 & 1 \\
Batch Size & $8\times64$ & $8\times64$ & 36 & 96 \\
GPUs & $8\times$A100 & $8\times$A100 & $3\times$A100 & $8\times$A100 \\
Audio Input Size & $128\times416$ & $128\times416$ & $16\times128\times416$ & $16\times128\times416$ \\
Visual Input Size & $224\times224$ & $224\times224$ & $16\times224\times224$ & $16\times224\times224$ \\
Patch Size (Audio/Visual) & $16\times16$ & $16\times16$ & $16\times16$ & $16\times16$ \\
Frames per Clip Used & 1 & 1 & 16 & 16 \\
MAE Masking Ratio & 0.75 & 0.75 & -- & -- \\
CL Masking Ratio & 0.50 & 0.50 & -- & -- \\
LTA Retention Ratio $\rho$ & -- & 0.50 & -- & -- \\
LTA Layer & -- & Last unified block & -- & -- \\
Encoder Frozen & No & No & Yes & Yes \\
Class-Balanced & No & No & No & Yes \\
Mixup & No & No & Yes & Yes \\
Random Shift & Yes & Yes & Yes & Yes \\
Loss Function & $\mathcal{L}_{\mathrm{base}}$ & $\mathcal{L}_{\mathrm{student}}$ & BCE & CE \\
Reconstruction Weight $\lambda_{\mathrm{rec}}$ & 1.0 & 1.0 & -- & -- \\
Contrastive Weight $\lambda_{\mathrm{cl}}$ & 0.1 & 0.1 & -- & -- \\
Contrastive Temperature $\tau$ & 0.05 & 0.05 & -- & -- \\
Anchor Weight $\lambda_a$ & -- & 0.5 & -- & -- \\
Consistency Weight $\lambda_c$ & -- & 0.5 & -- & -- \\
Weight Averaging & No & No & Yes & Yes \\
Norm Mean & $-5.081$ & $-5.081$ & $-5.081$ & $-5.081$ \\
Norm STD & $4.485$ & $4.485$ & $4.485$ & $4.485$ \\
\bottomrule
\end{tabular}
}
\vspace{0.3em}
\caption{Hyperparameters for pretraining and attention probing.
The MAE and CL masking ratios apply to the reconstruction and contrastive branches, respectively, in both modalities.
During student pretraining, LTA uses the frozen teacher's last unified block with retention ratio $\rho=0.50$ for both modalities.
The encoder-freezing setting refers to the encoder being optimized or probed in each column; the supervising teacher remains frozen during student training.
A dash denotes not applicable.}
\label{tab:pretrain-lp}
\end{table}

\subsection{Teacher Training Dynamics}
\label{app:teacher_dynamic}

To justify the use of an early-converged relevance teacher, we evaluate the quality of LTA-induced subviews at different teacher checkpoints.
For each checkpoint, we freeze the teacher, compute LTA scores, select the top-$\rho$ tokens with $\rho=0.50$, and feed the resulting subviews back into the teacher encoder.
We then measure zero-shot cross-modal retrieval performance using the subview representations, and report the average Recall@1 and Recall@5 over the two retrieval directions.

\begin{figure*}[t]
    \centering
    \includegraphics[width=\textwidth]{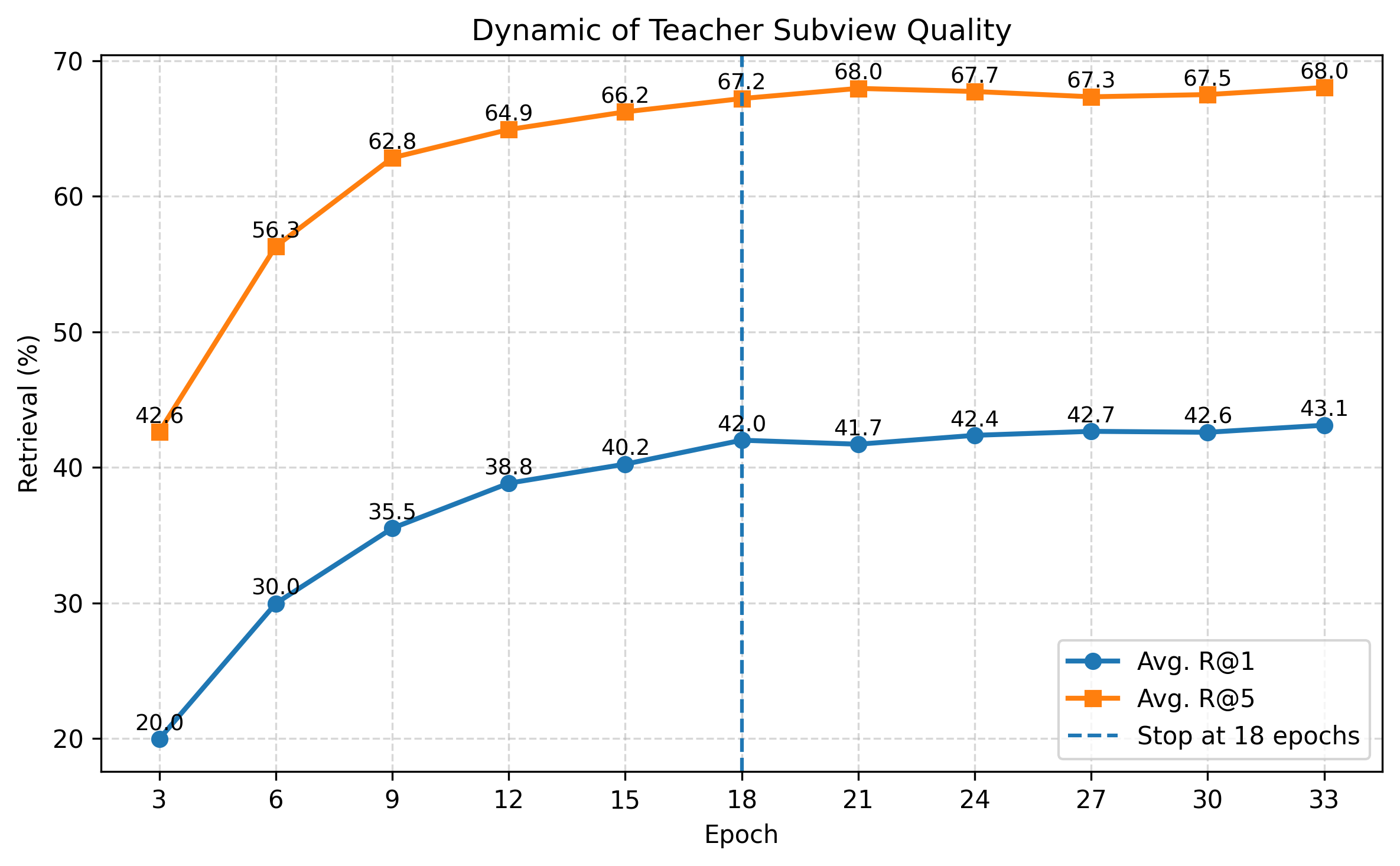}
    \caption{
    Dynamics of teacher-induced subview quality during teacher pretraining.
    For each teacher checkpoint, we construct LTA-induced subviews with retention ratio $\rho=0.50$ and evaluate zero-shot cross-modal retrieval using the resulting subview representations.
    Avg. R@1 and Avg. R@5 denote the mean retrieval performance over the two retrieval directions.
    The subview quality improves rapidly in the early stage and becomes saturated around epoch 18, after which later checkpoints provide only marginal gains.
    We therefore stop teacher pretraining at epoch 18 rather than using the full 35-epoch training schedule.
    }
    \label{fig:teacher_subview_dynamic}
\end{figure*}

As shown in Fig.~\ref{fig:teacher_subview_dynamic}, the quality of teacher-induced subviews improves rapidly during the early training stage.
The average R@1 increases from 20.0 at epoch 3 to 42.0 at epoch 18, while the average R@5 increases from 42.6 to 67.2.
After epoch 18, the retrieval performance becomes largely saturated: later checkpoints only bring marginal fluctuations, with average R@1 staying around 42--43 and average R@5 staying around 67--68.
This indicates that the token-relevance ranking used for subview construction has already become sufficiently stable by the middle stage of training.

The stopping epoch is selected using this teacher-side subview retrieval diagnostic.
We therefore stop teacher pretraining at epoch 18 instead of following the full 35-epoch TG-DP schedule.
This early stopping reduces the teacher-side training cost while preserving reliable subview supervision for student training.

\paragraph{Sensitivity to the teacher stopping epoch.}
We further evaluate how the teacher stopping epoch affects final student performance.
Using teachers stopped at epochs 15, 18, and 21, we train the corresponding students following the recipe in Appendix~\ref{app:training_details} and evaluate cross-modal retrieval on VGGSound and localization on AVSBench.
As shown in Table~\ref{tab:teacher_stop_epoch}, average retrieval R@1 remains within $34.50$--$34.86$, and the variation in either retrieval direction is at most $0.65$ R@1.
Localization performance ranges from $29.5$ to $30.6$ mIoU and from $36.5$ to $38.2$ mAP.
These results indicate that final student performance is relatively stable across the tested stopping epochs near the saturation stage, supporting the use of an early-stopped teacher.

\begin{table}[!htbp]
\centering
\small
\setlength{\tabcolsep}{5pt}
\renewcommand{\arraystretch}{1.08}
\begin{tabular*}{\linewidth}{@{\extracolsep{\fill}}cccccc@{}}
\toprule
\textbf{Teacher stop} & \multicolumn{3}{c}{\textbf{Retrieval}} & \multicolumn{2}{c}{\textbf{Localization}} \\
\cmidrule(lr){2-4}\cmidrule(lr){5-6}
\textbf{epoch} & V$\rightarrow$A R@1 & A$\rightarrow$V R@1 & Avg. R@1 & mIoU & mAP \\
\midrule
15 & 35.38 & 33.62 & 34.50 & 29.6 & 37.1 \\
\rowcolor{gray!12}
18 & 35.44 & 34.27 & 34.86 & 30.6 & 38.2 \\
21 & 35.19 & 33.81 & 34.50 & 29.5 & 36.5 \\
\bottomrule
\end{tabular*}
\caption{Sensitivity of final student performance to the teacher stopping epoch.
The default stopping epoch is shaded. Avg. R@1 is averaged over both retrieval directions.}
\label{tab:teacher_stop_epoch}
\end{table}

\FloatBarrier
\Needspace{0.52\textheight}
\subsection{Repeated Runs}
\label{app:repeated_runs}

To assess the variability of the reported results, we conduct five independent pretraining runs with different random seeds and evaluate the resulting models on VGGSound retrieval and AVSBench localization.
For AS20K classification, we repeat attention probing five times using the same pretrained checkpoint with the encoders frozen.
The classification results therefore measure variability in downstream probing, while the retrieval and localization results reflect variability across independent pretraining runs.
Table~\ref{tab:repeated_runs} reports the mean and standard deviation for each metric.

The standard deviations of VGGSound retrieval R@1 are $0.35$ and $0.70$ for V$\rightarrow$A and A$\rightarrow$V, respectively.
On AVSBench, the standard deviations are $0.44$ mIoU and $0.57$ mAP, while repeated AS20K probing yields a standard deviation of $0.13$ mAP.
These results support the stability of the reported performance under the tested training and evaluation settings.

\begin{table}[!htbp]
\centering
\small
\setlength{\tabcolsep}{3pt}
\renewcommand{\arraystretch}{1.08}
\begin{tabular}{@{}lccc@{}}
\toprule
& \multicolumn{2}{c}{\textbf{AVSBench localization}}
& \textbf{AS20K classification} \\
\cmidrule(lr){2-3}\cmidrule(lr){4-4}
\textbf{Method} & mIoU & mAP & mAP \\
\midrule
Ours & $29.88 \pm 0.44$ & $37.26 \pm 0.57$ & $35.20 \pm 0.13$ \\
\bottomrule
\end{tabular}
\par\smallskip
\textbf{(a) Localization and classification}\par

\medskip
\begin{tabular}{@{}lcccccc@{}}
\toprule
& \multicolumn{3}{c}{\textbf{Vision $\rightarrow$ Audio}}
& \multicolumn{3}{c}{\textbf{Audio $\rightarrow$ Vision}} \\
\cmidrule(lr){2-4}\cmidrule(lr){5-7}
\textbf{Method} & R@1 & R@5 & R@10 & R@1 & R@5 & R@10 \\
\midrule
Ours & $35.32 \pm 0.35$ & $59.76 \pm 0.64$ & $69.54 \pm 0.63$
     & $33.95 \pm 0.70$ & $58.93 \pm 0.82$ & $69.51 \pm 0.30$ \\
\bottomrule
\end{tabular}
\par\smallskip
\textbf{(b) VGGSound retrieval}\par
\caption{Repeated-run results, reported as mean $\pm$ standard deviation.
Retrieval and localization results use five independent pretraining runs with different random seeds.
Classification results use five attention-probing runs with the same pretrained checkpoint.
The corresponding means in Tables~\ref{tab:retrieval}--\ref{tab:segmentation} are rounded to one decimal place.}
\label{tab:repeated_runs}
\end{table}
\FloatBarrier
\section{Additional Analysis of Layer-Truncated Attribution}
\label{app:lta_analysis}
\subsection{Computation Cost}
\label{app:lta_cost}
We report the overall pretraining time and the per-pair computation cost of LTA scoring separately.
The teacher, attribution computation, and subview construction are used only during pretraining and introduce no additional inference-time overhead.

\paragraph{Overall pretraining cost.}
Under the same hardware setting of $8$ NVIDIA A100 GPUs, the adopted TG-DP backbone requires $19.0$ hours for $35$ epochs of pretraining.
Our method first trains a teacher for $18$ epochs and then trains the student for $25$ epochs, following Appendix~\ref{app:training_details}.
Table~\ref{tab:training_cost} provides the corresponding breakdown in wall-clock hours.
Teacher pretraining takes $9.77$ hours, while the student stage consists of $13.57$ hours of base computation, $2.40$ hours of additional consistency regularization, and $1.39$ hours of teacher-anchor construction and LTA scoring.
These components sum to $27.13$ hours, reported as $27.1$ hours in the main text.
The main source of additional training cost is the one-time pretraining of the early-stopped teacher.
\begin{table}[!htbp]
\centering
\small
\setlength{\tabcolsep}{6pt}
\renewcommand{\arraystretch}{1.05}
\begin{tabular}{@{}lr@{}}
\toprule
\textbf{Training component} & \textbf{Time (hours)} \\
\midrule
TG-DP baseline ($35$ epochs) & 19.00 \\
\midrule
Teacher pretraining ($18$ epochs) & 9.77 \\
Student base computation ($25$ epochs) & 13.57 \\
Additional consistency regularization & 2.40 \\
Teacher-anchor construction and LTA scoring & 1.39 \\
\midrule
\textbf{Our total pretraining time} & \textbf{27.13} \\
\bottomrule
\end{tabular}
\caption{Breakdown of pretraining time under the same hardware setting of $8$ NVIDIA A100 GPUs. Times are reported in wall-clock hours. The four components below the baseline sum to our total pretraining time; the last three comprise the student stage.}
\label{tab:training_cost}
\end{table}

\paragraph{Per-pair cost of LTA scoring.}
To isolate the model computation involved in LTA scoring, we exclude data loading and audio/image preprocessing and report the cost per audio-visual pair.
As shown in Table~\ref{tab:lta_cost}, a standard teacher forward pass costs $71.78$ GFLOPs, while the teacher forward together with LTA scoring costs $76.47$ GFLOPs.
The additional attribution computation therefore requires $4.69$ GFLOPs, or $6.54\%$ of the teacher forward cost.
Memory usage increases from $746.83$ MB to $779.10$ MB, an increment of $32.27$ MB.
The $6.54\%$ figure describes the incremental FLOPs of attribution relative to a teacher forward pass; it does not represent the increase in total pretraining time.
Likewise, the $1.39$-hour component in Table~\ref{tab:training_cost} includes both teacher-anchor construction and LTA scoring.
\begin{table}[!htbp]
\centering
\small
\setlength{\tabcolsep}{5pt}
\renewcommand{\arraystretch}{1.05}
\begin{tabular}{@{}lrrr@{}}
\toprule
\textbf{Component} & \textbf{GFLOPs} & \textbf{Memory (MB)} & \textbf{Relative FLOPs} \\
\midrule
Teacher forward & 71.78 & 746.83 & 1.00$\times$ \\
Teacher forward + LTA & 76.47 & 779.10 & 1.07$\times$ \\
LTA overhead & +4.69 & +32.27 & +6.54\% \\
\bottomrule
\end{tabular}
\caption{Computation cost of LTA scoring per audio-visual pair, excluding data loading and audio/image preprocessing. The full scoring process includes one teacher forward pass and the additional attribution computation. Relative FLOPs are computed with respect to the teacher forward.}
\label{tab:lta_cost}
\end{table}
\FloatBarrier

\subsection{Segmentation Ability of LTA}
\label{app:lta_segmentation}
\begin{table}[t]
\centering
\setlength{\tabcolsep}{8pt}
\renewcommand{\arraystretch}{1.08}
\small
\begin{tabular}{@{}lcccc@{}}
\toprule
& \multicolumn{2}{c}{\textbf{AVSBench}} 
& \multicolumn{2}{c}{\textbf{ADE20K-Sound}} \\
\cmidrule(lr){2-3}\cmidrule(lr){4-5}
\textbf{Method} 
& \textbf{mIoU} & \textbf{mAP} 
& \textbf{mIoU} & \textbf{mAP} \\
\midrule
Original heatmap & 22.3 & 25.2 & 21.9 & 25.4 \\
CheferCAM        & \textbf{44.5} & \textbf{61.2} & \textbf{32.3} & \textbf{41.9} \\
\rowcolor{gray!12}
LTA              & 44.2 & 60.1 & 31.1 & 39.6 \\
\bottomrule
\end{tabular}
\vspace{0.3em}
\caption{
Sound-prompted semantic segmentation results of different heatmaps on AVSBench and ADE20K-Sound.
We report mean IoU (mIoU) and mean average precision (mAP) in percentage.
}
\label{tab:lta_seg}
\end{table}
We further evaluate whether LTA captures fine-grained audio-visual correspondence beyond global retrieval.
To this end, we use sound-prompted semantic segmentation as an external localization test.
Given an audio prompt and a visual image or video frame, the model produces a spatial heatmap indicating regions that are relevant to the input sound.
We compare three types of heatmaps: the original audio-visual similarity heatmap, the full CheferCAM attribution heatmap, and our LTA heatmap.

As shown in Table~\ref{tab:lta_seg}, attribution-based heatmaps substantially improve over the original similarity heatmap on both ADE20K-Sound and AVSBench.
On ADE20K-Sound, LTA improves mIoU from 21.9 to 31.1 and mAP from 25.4 to 39.6.
On AVSBench, LTA improves mIoU from 22.3 to 44.2 and mAP from 25.2 to 60.1.
These results indicate that gradient-attention attribution provides a much stronger localization signal than raw cross-modal similarity.

Compared with full CheferCAM, LTA achieves very similar segmentation performance.
On AVSBench, the gap is only 0.3 in mIoU and 1.1 in mAP, showing that last-layer gradient-attention relevance already preserves most of the localization ability needed for identifying sound-related regions.
Although CheferCAM performs full relevance propagation through the Transformer, its marginal gain is small for this subview-selection purpose.
Therefore, LTA offers a favorable trade-off: it retains strong fine-grained localization ability while avoiding the computational cost of full-network attribution propagation.
This supports our choice of LTA as an efficient token-relevance estimator for constructing shared-cover subviews.

\subsection{Visualization of LTA}
\label{app:lta_vis}
\begin{figure*}[t]
    \centering
    \includegraphics[width=\textwidth]{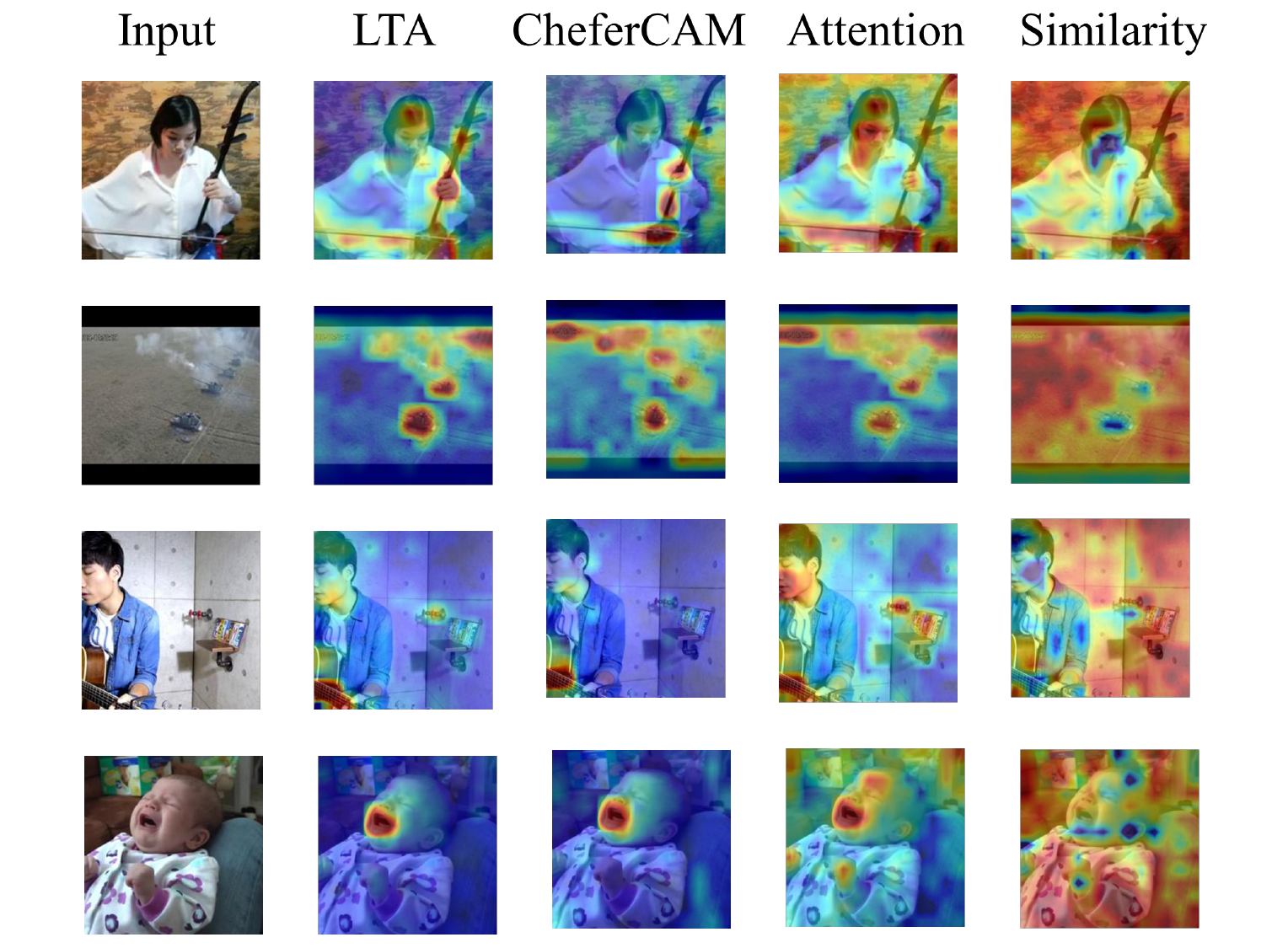}
    \caption{
    Visualization of different score type.
}
    \label{fig:lta_visualization}
\end{figure*}
We provide qualitative visualizations to further examine whether LTA captures sound-relevant regions at the token level.
Fig.~\ref{fig:lta_visualization} compares heatmaps generated by different scoring functions, including LTA, CheferCAM, raw attention, and token-level similarity.
For each example, warmer regions indicate tokens assigned with higher relevance scores.

The visualization shows that LTA produces compact and semantically meaningful heatmaps around sound-related objects or regions.
For example, it highlights the musical instrument in the first row, the sounding object region in the second row, the object being interacted with in the third row, and the mouth/face region of the crying infant in the fourth row.
These regions are consistent with the expected audio-visual correspondence in each sample.
CheferCAM produces similar localization patterns, but it requires full relevance propagation through the Transformer, which introduces substantial computational overhead. 
The close visual agreement between LTA and CheferCAM suggests that last-layer gradient-attention relevance preserves the core attribution behavior needed for subview selection, while avoiding expensive full-network propagation.

In contrast, raw attention and similarity-based scores are less reliable.
Attention maps often spread over visually salient but acoustically irrelevant areas, since attention weights only reflect forward token aggregation and are not explicitly conditioned on the cross-modal matching objective.
Similarity maps can also be noisy or biased toward broad appearance-level correlations, leading to less precise localization.
These qualitative results are consistent with the quantitative segmentation results in Table~\ref{tab:lta_seg}, and further support the use of LTA as an efficient token-level scoring function for constructing shared-cover subviews.
\section{Limitations}
\label{app:limitations}

Our work has several limitations. First, although the motivation of shared-private
decoupling is generally applicable to multimodal representation learning, our
experiments are focused on the audio-visual setting. Due to the high cost of
large-scale multimodal pretraining, we have not yet validated the proposed
framework on other modality pairs, such as image-text or audio-language data.
Extending attribution-guided shared-private decoupling to broader multimodal
settings is an important direction for future work.

Second, the decoupling ability internalized by the student is still limited.
As shown in Table~\ref{tab:decouple}, the private token still retains a
non-negligible amount of cross-modal shared information, indicating that the
learned shared and private representations are not completely separated.
Moreover, the retrieval performance of the student shared token does not fully
match the teacher-induced shared-cover subview in
Table~\ref{tab:lta_subview_retrieval}. This suggests that the current
teacher-student training transfers the attribution-induced decoupling behavior
only partially. More effective distillation objectives or architectural designs
may further improve the separation between shared and private representations.

Third, LTA is a heuristic approximation to the ideal shared-relevance score used
in our theoretical analysis. The theory assumes a ranking function that reflects
each token's contribution to the cross-modal shared factor, while LTA estimates
this relevance using the last-layer gradient-attention signal of the matching
objective. Although our empirical results show that LTA provides an effective
ranking signal for subview construction, it does not provide a strict guarantee
of recovering the theoretically ideal score. Developing more principled and
theoretically grounded token-level relevance estimators remains an interesting
future direction.

\end{document}